# FLOW-Methode

## - Methodenbeschreibung zur Anwendung von FLOW -

Kai Stapel und Kurt Schneider

27.02.2012

Version 1.5

Arbeitspaket 6 (AP6) „Konsolidierung und Dokumentation der Methode“

Projekt InfoFLOW, 2008-2011, gefördert durch die DFG

**Fachgebiet Software Engineering**

Leibniz Universität Hannover

Welfengarten 1, 30167 Hannover

Kai Stapel

Telefon: +49-(0)511-762-7463

kai.stapel@inf.uni-hannover.de

www.se.uni-hannover.de

Prof. Dr. Kurt Schneider

Telefon: +49-(0)511-762-19666

kurt.schneider@inf.uni-hannover.de

# Inhaltsverzeichnis

# Executive Summary

In Softwareprojekten und in Software-Unternehmen fließen vielerlei Informationen. Anforderungen müssen vom Kunden zu den Entwicklern gelangen. Auch Tester müssen die Anforderungen kennen. Randbedingungen und Entwurfsentscheidungen müssen zur rechten Zeit am rechten Ort sein.

Um dies zu erreichen, wurden Prozesse und Workflows entwickelt. Sie sollen alle Beteiligten anleiten, die richtigen Tätigkeiten auszuführen und bestimmte Dokumente zu erstellen. Doch inzwischen ist diese rein dokumentenorientierte Sicht überholt: In agilen oder leicht-gewichtigen Entwicklungsmethoden wird direkte Kommunikation sehr wichtig genommen. Nicht alles kann oder soll niedergeschrieben werden. Viele Unternehmen brauchen auch solche Elemente in ihrer Softwareentwicklung.

Die Informationsflussanalyse mit FLOW ermöglicht es, unabhängig von der Entwicklungsmethode zu modellieren, wie und auf welchem Wege Informationen und Erfahrungen fließen. Erfahrungen spielen dabei oft die Rolle von Steuergrößen, denn erfahrene Mitarbeiter können Informationen kompetenter bearbeiten und weiterleiten. Auch die Erfahrungen müssen in geeigneter Form zur rechten Zeit am rechten Ort sein. Viele Erfahrungen werden aber nie dokumentiert. Da Informationen und Erfahrungen sowohl in agilen als auch in traditionellen Umgebungen fließen müssen, wird in FLOW ein Modell aufgebaut, das nicht nach agil, traditionell oder anderen Bezeichnungen unterscheidet, sondern einzig danach, wie die Flüsse gestaltet sind.

Kernkonzepte und Alleinstellungsmerkmale der Modellierung und Verbesserung von Informationsflüsse in FLOW sind:

- Modellierung von festen (dokumentierten) und flüssigen (im Wesentlichen: nicht dokumentierten) Informationsflüssen. Beide haben Vor- und Nachteile. Es wird nicht davon ausgegangen, dass alle Informationen fest und dokumentiert sein müssten.
- Berücksichtigung von Erfahrungen als einer wesentlichen Sorte von Informationen. Sie werden explizit mit modelliert. Häufig sind Erfahrungen flüssig, also an erfahrene Personen gebunden, wodurch sie in der Prozessmodellierung leicht aus der Betrachtung fallen.

Dieses Dokument gibt eine Übersicht über die Schritte der Methode FLOW zur Informationsflussmodellierung und -optimierung. Es wird gezeigt, welche Ziele man dabei verfolgen kann und was dabei zu beachten ist. Die Schritte der wichtigsten Aktivitäten werden vorgestellt.

So einfach die Grundkonzepte aussehen, so kompliziert und vielfältig können die Konsequenzen aus einer erkannten Informationsflussanomalie sein. Im einfachsten Fall soll eine flüssige Information "verfestigt" werden, also in geeigneter Form dokumentiert. Selten reicht es aus, jemanden dazu aufzufordern, etwas aufzuschreiben, was er oder sie sonst im Kopf behalten würde. Zeitmangel und andere Prioritäten haben eben gerade dazu geführt, die entsprechenden Informationen lieber flüssig im Kopf zu behalten. Dort werden sie leider leicht vergessen. Flüssige Informationen zerrinnen eben auch leicht, um in der Metapher zu bleiben. Dagegen dauert es länger und kostet mehr Zeit, Informationen fest zu halten, sie also zu dokumentieren. Es ist meist auch aufwändiger, ein Dokument zu lesen, als jemanden zu fragen. In der FLOW-Methode geht es im Wesentlichen darum, die individuell beste Balance von festen und flüssigen Informationen zu finden. Dazu gehört, den Betroffenen die "Verfestigung" so einfach wie möglich zu machen.

Im Laufe der letzten Jahre wurde eine ganze Reihe spezialisierter Techniken entwickelt, um unter bestimmten Umständen genau die gewünschte Verbesserung von Flüssen, Aggregatzuständen (fest und flüssig) und Erfahrungsweitergabe zu bewerkstelligen. Jede solche Technik kann durch Werkzeuge unterstützt sein und berücksichtigt in der Regel die kognitiven Möglichkeiten und Grenzen der handelnden Personen.

Neben der Methodenbeschreibung bietet dieses Dokument auch eine Übersicht über etliche Techniken. Dies illustriert einerseits, wie solche Techniken beschaffen sind. Zum anderen können die angeführten Beispiele konkret genutzt werden, um in anderen Unternehmen und Situationen ähnliche Anliegen zu bearbeiten. Die präsentierten Techniken stellen also ein erstes Repertoire möglicher Verbesserungsmaßnahmen dar, die weit über einfache Fest-flüssig-Umwandlungen hinausgehen. Die Entscheidung, für eine gegebene Situation die geeignetste Technik zu finden, wird durch einen Profilbogen unterstützt, der ebenfalls vorgestellt wird.

Das Forschungsprojekt InfoFLOW läuft seit drei Jahren. Schon zuvor wurden Vorarbeiten auf der Basis industrieller Erfahrungen geleistet und in Verbesserungsideen übersetzt. Die FLOW-Methode wurde in der Zwischenzeit auf viele Forschungssituationen und in mehreren Unternehmen zur konkreten Informationsflussverbesserung angesetzt. FLOW ist aber erst am Beginn der Operationalisierung als gefestigte Beratungs- und Verbesserungsmethode. Es gibt noch viele Informationen und Erfahrungen, die nirgends dokumentiert sind. Noch kommen bei jedem Einsatz neue Ideen hinzu, nur der Kern der Methode hat sich inzwischen stabilisiert. Der Zweck des vorliegenden Dokuments ist es, diesen Kern zu "verfestigen". Wir betrachten diese Methodenbeschreibung als den ersten Schritt, den Einsatz von FLOW auch schriftlich zu erklären.

# 1 Einleitung

Die FLOW-Methode ist eine Methode zur Analyse und Verbesserung von Informationsflüssen in Softwareprojekten. Eine Besonderheit der FLOW-Methode ist, dass nicht nur dokumentierte Informationen, sondern auch die Informationen in die Betrachtung eingeschlossen werden, die mündlich weiter gegeben werden. Damit eignet sich die FLOW-Methode sowohl zur Analyse und Verbesserung von dokumentenzentrierten Prozessen als auch zur Optimierung agiler kommunikationsintensiver Entwicklung.

Die FLOW-Methode kann z.B. von Projektleitern eingesetzt werden, um ein laufendes Projekt zu verbessern oder Informationsflüsse für zukünftige Projekte gezielt zu planen. Eine andere Anwendungsmöglichkeit der Methode ist der Einsatz durch Prozessabteilungen zur organisationsweiten projektübergreifenden Softwareprozessverbesserung. Weiterhin kann die FLOW-Methode auch im Kleinen, zur Verbesserung einzelner Entwicklungsaktivitäten genutzt werden. Welchen Umfang ein FLOW-Verbesserungsvorhaben haben soll wird in der Vorbereitungsphase festgelegt (vgl. Kapitel 3.2).

Das Ziel dieser Methodenbeschreibung ist die praktische Nutzbarmachung der im InfoFLOW-Projekt erarbeiteten theoretischen Grundlagen zur Analyse und Verbesserung von Informationsflüssen in Softwareprojekten. Damit der Anwender der FLOW-Methode die Techniken besser verstehen und ggf. auch eigene Techniken entwickeln kann, werden in diesem Dokument auch die FLOW-Grundlagen vorgestellt (vgl. Kapitel 2). Die FLOW-Methode selbst wird in Kapitel 3 beschrieben. Konkrete praktische Anleitungen zur Anwendung der FLOW-Methode werden durch FLOW-Techniken beschrieben. Dieses Dokument enthält eine Übersicht über elf FLOW-Techniken und jeweils Verweise auf weiterführende Informationen (vgl. Kapitel 4).

## 1.1 Für wen ist die FLOW-Methode?

Die FLOW-Methode kann von

- externen FLOW-Experten,
- Mitarbeitern interner Prozessabteilungen,
- der Projektleitung oder
- individuell von einzelnen Softwareentwicklern

eingesetzt werden, um Informationsflüsse einzelner Entwicklungsaktivitäten, von Softwareprojekten oder der gesamten Organisation zu analysieren oder zu verbessern.

## 1.2 Vorbedingungen

Damit die FLOW-Methode erfolgreich eingesetzt werden kann, müssen einige Vorbedingungen erfüllt sein. Vorbedingungen für die Durchführung FLOW-basierter Softwareprozessverbesserung sind:

- Das zu verbessernde Projekt (bzw. die Aktivität oder die Organisation) muss bereit sein, ihre Informationsflüsse für die Analyse preiszugeben.
- Das analysierte Projekt (Aktivität, Organisation) muss offen für Verbesserungsvorschläge sein und Änderungen flexibel umsetzen können und wollen.
- Es müssen für die Problemstellung (FLOW-Ziel und Projektparameter) passende FLOW-Techniken inkl. zugehöriger Werkzeuge zur Verfügung stehen (vgl. Kapitel 4).

## 1.3 Aufbau

Diese Methodenbeschreibung ist wie folgt aufgebaut:

In *Kapitel 2* werden FLOW-Grundlagen des hinter der Methode stehenden Forschungsprojekts InfoFLOW beschrieben. Die Grundkonzepte werden erläutert. Die Metapher der Aggregatzustände wird eingeführt. Sie ist das wichtigste Element der Methode. Abschließend wird die grafische FLOW-Notation zur Modellierung von Informationsflussmodellen vorgestellt.

In *Kapitel 3* wird die *FLOW-Methode* selbst beschrieben. Zunächst wird auf die einzelnen Ziele, die mit der Methode verfolgt werden (können), eingegangen, um anschließend die drei Phasen der Methode detailliert beschreiben zu können. Die Methode ist die theoretische Grundlage für alle Techniken, die im Projekt InfoFLOW entstanden sind und weiterhin entstehen.

In *Kapitel 4* werden *FLOW-Techniken* beschrieben, die im Projekt InfoFLOW entstanden sind, um die Methode praktisch anzuwenden. Jede Phase hat spezielle Techniken, die teilweise durch manuelle oder elektronische Werkzeuge unterstützt werden können. Die Techniken beschreiben die praktische Anwendung der Methode. Ein konkretes Verbesserungsvorhaben wird durch die Anwendung mindestens einer FLOW-Technik umgesetzt.

Im Anhang befinden sich zusätzliche Materialien, die u. a. zur Durchführung der verschiedenen Techniken notwendig sind.

# 2 FLOW-Grundlagen

Bei der Softwareentwicklung gibt es trotz reifer Prozesse nach wie vor Probleme. Projekte verspäten sich oder schlagen ganz fehl (vgl. z.B. [3]). Vorgeschriebene Dokumente werden wegen hohem Zeitdruck nicht erstellt. Der vorgegebene Prozess wird nicht eingehalten. Die vielversprechenden agilen Methoden lassen sich nicht oder nur kaum mit den schwergewichtigen Prozessen vereinbaren. Was kann man tun, um die Vorteile beider Seiten (dokumentenzentriert und agil) zu nutzen?

Information ist die Basis aller Entwicklungsaktivitäten. Denn egal wie entwickelt wird, ob mit einem dokumentenzentrierten Prozess oder agil, die Informationen, die der Kunde in Form von Anforderungen im Kopf hat, müssen möglichst vollständig und möglichst schnell in das Endprodukt Software gelangen. Daher eignen sich Informationen und deren Fluss im Softwareentwicklungsprojekt als Ausgangspunkt für die Analyse und Verbesserung von Softwareprojekten. Das Hauptziel der FLOW-Methode ist daher die Verbesserung von Informationsflüssen in der Softwareentwicklung.

Die Analyseverfahren von FLOW beruhen auf einigen Grundannahmen, die im Folgenden beschrieben werden.

## 2.1 Konzepte

Die FLOW-Forschung basiert auf wenigen Grundannahmen, die sich aus dem Hauptziel abgeleitet und in der industriellen Praxis als sinnvoll herausgestellt haben:

- Informationsflüsse sind Bindeglied zwischen dokumentenlastigen (z. B. V-Modell XT [21] oder RUP [4]) und kommunikationsintensiven (z. B. agil, eXtreme Programming [1], SCRUM [16]) Ansätzen.
- Modelle dienen in erster Linie als Diskussionsgrundlage und sind daher so einfach wie möglich, d.h. aber auch so detailliert wie nötig.
- Inhalte werden meist nur grob modelliert.
- Erfahrungen sind eine besonders wertvolle Art von Informationen (vgl. [13]).
- Informationen haben einen Aggregatzustand.

## 2.2 Metapher der Aggregatzustände

Eine wesentliche Eigenschaft von FLOW ist die Unterscheidung von **festen** und **flüssigen** Informationen und abgeleiteten Flüssen. Die Metapher ist darauf ausgerichtet schnell zwischen wesentlichen Eigenschaften von Informationsflüssen in verschiedenen Medien unterscheiden zu können. **Feste** Informationen sind für diejenigen, die sie benötigen, unverändert wiederholt zugreifbar und verständlich, und das auch noch nach längerer Zeit. **Flüssige** Informationen hingegen sind nach einer gewissen Zeit nicht mehr abrufbar, z.B. wenn sie vergessen wurden. Flüssige Informationen haben aber den Vorteil, dass sie leichter, d.h. schneller, weitergegeben werden können, weil bei ihnen nicht darauf geachtet werden muss, dass sie jeder verstehen und abrufen kann. Feste und flüssige Informationen sind daher wie folgt definiert:

*Definition:* **Feste Informationen** sind Informationen, die in einem Betrachtungsbereich von allen jederzeit *abgerufen* und *verstanden* werden können.

*Definition:* **Flüssige Informationen** sind Informationen, die nicht fest sind.

Die Unterscheidung zwischen fest und flüssig hängt vom Betrachtungsbereich ab. Ein Betrachtungsbereich ist wie folgt definiert.

*Definition:* Ein Betrachtungsbereich ist

- eine Menge von Personen und
- eine Zeitspanne.

D.h. Informationen sind in einem Betrachtungsbereich dann fest, wenn sie

- von allen Personen in diesem Betrachtungsbereich und
- zu jedem Zeitpunkt innerhalb der Zeitspanne des Betrachtungsbereichs

abgerufen und verstanden werden können.

Aus diesen Definitionen ergeben sich bestimmte Eigenschaften für Informationen in den beiden Aggregatzuständen. Tabelle 1 listet diese Eigenschaften mit einigen typischen Beispielen auf.

**Tabelle 1. Eigenschaften und Beispiele der Aggregatzustände *Fest* und *Flüssig*.**

| Aggregatzustand | Eigenschaften | Beispiele |
|---|---|---|
| **Fest** | Vorteile<br>• Wiederholt abrufbar<br>• Langfristig zugänglich<br>• Dritten zugänglich<br>• Dritte können etwas damit anfangen<br>Nachteile<br>• Speicherung kostet Zeit und Aufwand<br>• Abruf kostet Zeit und Aufwand | • Dokumente (analog und elektronisch)<br>• Audio-/Video-Aufnahmen<br>• Screencasts<br>• Quellcode |
| **Flüssig** | Vorteile<br>• Können leicht (effektiv und effizient) gespeichert und weitergegeben werde<br>• Können leicht (effektiv und effizient) abgerufen und verstanden werden, wenn der notwendige Kontext vorhanden ist<br>Nachteile<br>• Gehen leicht verloren (Information selbst oder Kontext wird vergessen), sie "versickern".<br>• Können leicht missverstanden werden, wenn der notwendige Kontext fehlt | • Gedächtnis von Personen<br>• informelle E-Mails<br>• Chat-Protokolle<br>• Notizzettel<br>• Whiteboard-Notizen |

## 2.3 Grafische Notation

Ausgehend von den Grundkonzepten und der Metapher der Aggregatzustände wurde eine Notation entwickelt, die es erlaubt, die wesentlichen Eigenschaften von FLOW-Modellen, nämlich die explizite Unterscheidung von *festen* und *flüssigen* Informationsflüssen, intuitiv darstellen zu können. Tabelle 2 gibt einen Überblick über die Syntax der FLOW-Notation. Im Wesentlichen gibt es je ein Symbol für feste und flüssige Informationsspeicher und entsprechend je einen Pfeiltyp für Informationsflüsse. Für den Fall, dass man noch nicht weiß (Ist-Modell) oder noch nicht festgelegt hat (Soll-Modell), welchen Typ ein Fluss hat, gibt es undefinierte bzw. unbekannte Speicher und Flüsse. Ergänzt wird das Ganze durch die Möglichkeit über das Aktivitäten-Symbol FLOW-Modelle zu hierarchisieren oder an bestehende Prozess-Notationen, wie EPKs oder UML-Aktivitätsdiagramme, zu koppeln.

### Fest

Das Symbol für einen *festen Informationsspeicher* ist das Dokument. Es wurde gewählt, weil es in der Softwareentwicklung der typischste Vertreter von festen Informationen ist. Möchte man als Informationsspeicher mehrere Dokumente eines Typs modellieren, so stellt man das mit dem 3-fach hinterlegten Dokumentensymbol dar. Alle Informationsflüsse, die aus einem festen Speicher hervorgehen, sind wiederum fest. Sie werden mit einem Pfeil mit durchgehender Linie dargestellt.

### Flüssig

Das Symbol für einen *flüssigen Informationsspeicher* ist ein Smiley. Es wurde gewählt, weil flüssige Informationen immer an Personen gebunden sind. Eine Gruppe von Personen wird mit dem 3-fachen Smiley dargestellt. Falls die Unterscheidung zwischen Rollen und Individuen bei Personen erforderlich ist, kann man dies durch verschiedene Bezeichner darstellen. So kann zum Beispiel durch Unter-streichen des Namens verdeutlicht werden, dass es sich um eine Rolle handelt (z. B.: „Müller“ im Vergleich zu „Analyst“). Informationsflüsse, die von einem flüssigen Speicher ausgehen, sind flüssig. Sie werden mit einem Pfeil mit gestrichelter Linie dargestellt.

## Undefinierter / unbekannter Aggregatzustand

Das Symbol für einen *undefinierten bzw. unbekannten Informationsspeicher* ist eine Kombination aus den Symbolen für flüssige und feste Speicher. Eine Gruppe von undefinierten / unbekannten Informationsspeichern wird mit dem 3-fachen Symbol dargestellt. Alle Informationsflüsse, die aus einem undefinierten / unbekannten Speicher hervorgehen, sind wiederum undefiniert / unbekannt.

Die Bedeutung dieses Aggregatzustands variiert je nach Typ des FLOW-Modells. In Soll-Modellen steht er für Speicher und Flüsse, bei denen nicht festgelegt wurde, welcher Aggregatzustand dargestellt wird. Dieser Aggregatzustand steht daher in *Soll*-Modellen für *undefiniert*. In Ist-Modellen steht er für Speicher und Flüsse, bei denen nicht bekannt ist, welcher Aggregatzustand vorliegt. Dieser Aggregatzustand steht daher in *Ist*-Modellen für *unbekannt*.

## Erfahrung

Da *Erfahrung* eine in der Softwareentwicklung besonders wertvolle Art von Information ist, kann man sie in FLOW explizit darstellen. Dies geschieht durch eine andere Farbe der Speicher und Flüsse. Meist wird Grau als Kontrast zu den sonst schwarzen Symbolen und Bezeichnern verwendet, so dass eine Unterscheidung auch noch auf Schwarz-Weiß-Ausdrucken möglich ist.

**Tabelle 2. Syntax der grafischen FLOW Notation**

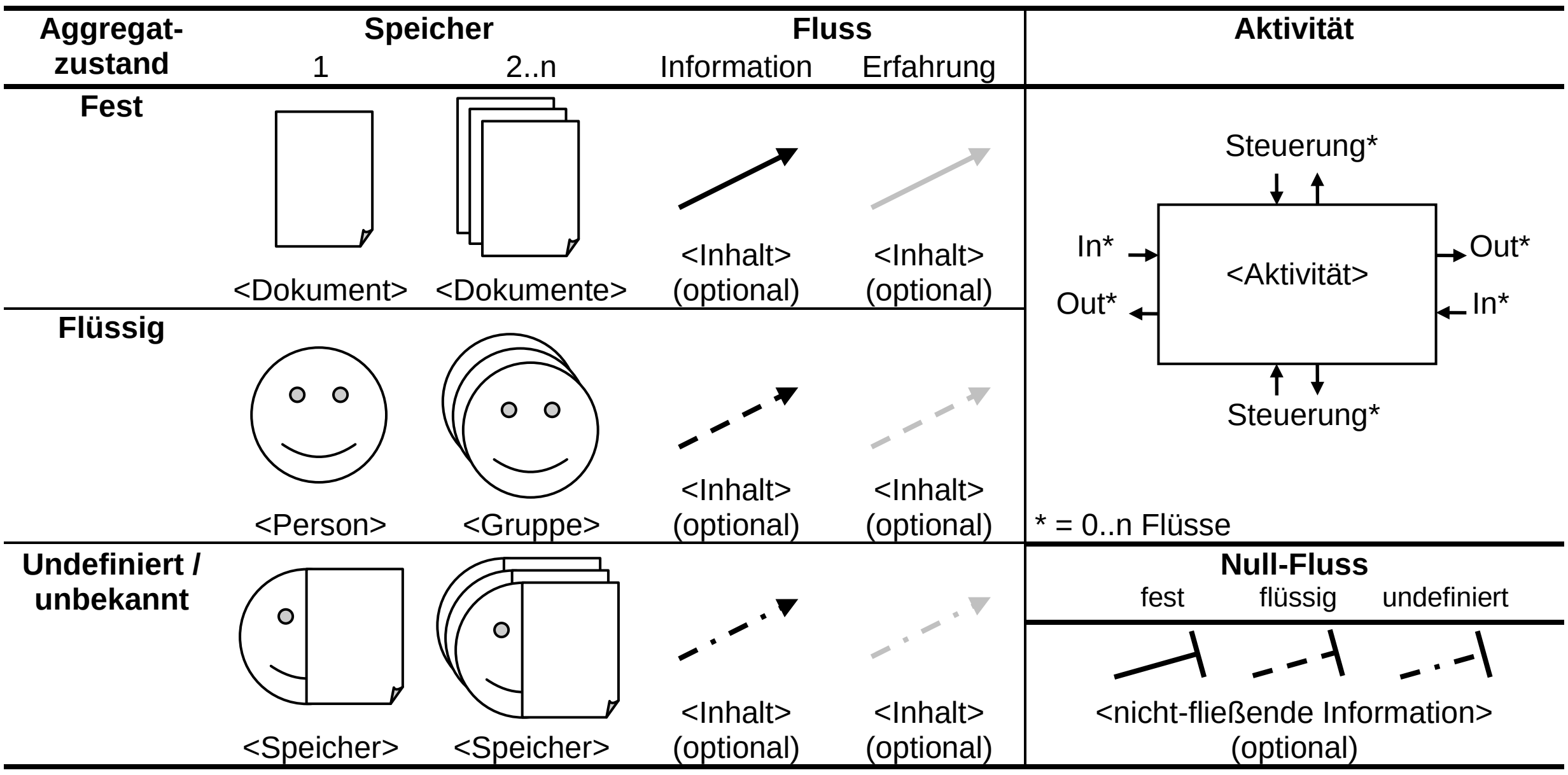

## Aktivität

In FLOW-Modellen hat das Aktivitäts-Symbol zwei Aufgaben:

1. **Hierarchisierung:** Es fasst Informationsspeicher (und ggf. andere Aktivitäten) zusammen und verbirgt sie. Damit ermöglicht es eine hierarchische Strukturierung der Modelle. Die eingehenden und ausgehenden Flüsse einer Aktivität, das so genannte FLOW-Interface, müssen konsistent mit den ein- und ausgehenden Flüssen im darunter liegenden Detailmodell sein.
2. **Prozessanbindung:** Es dient als Verbindung zu bestehenden Prozessnotationen. Aspekte von Informationsflüssen, insbesondere die Unterscheidung zwischen fest und flüssig, können so in bestehende Prozessdarstellungen integriert werden.

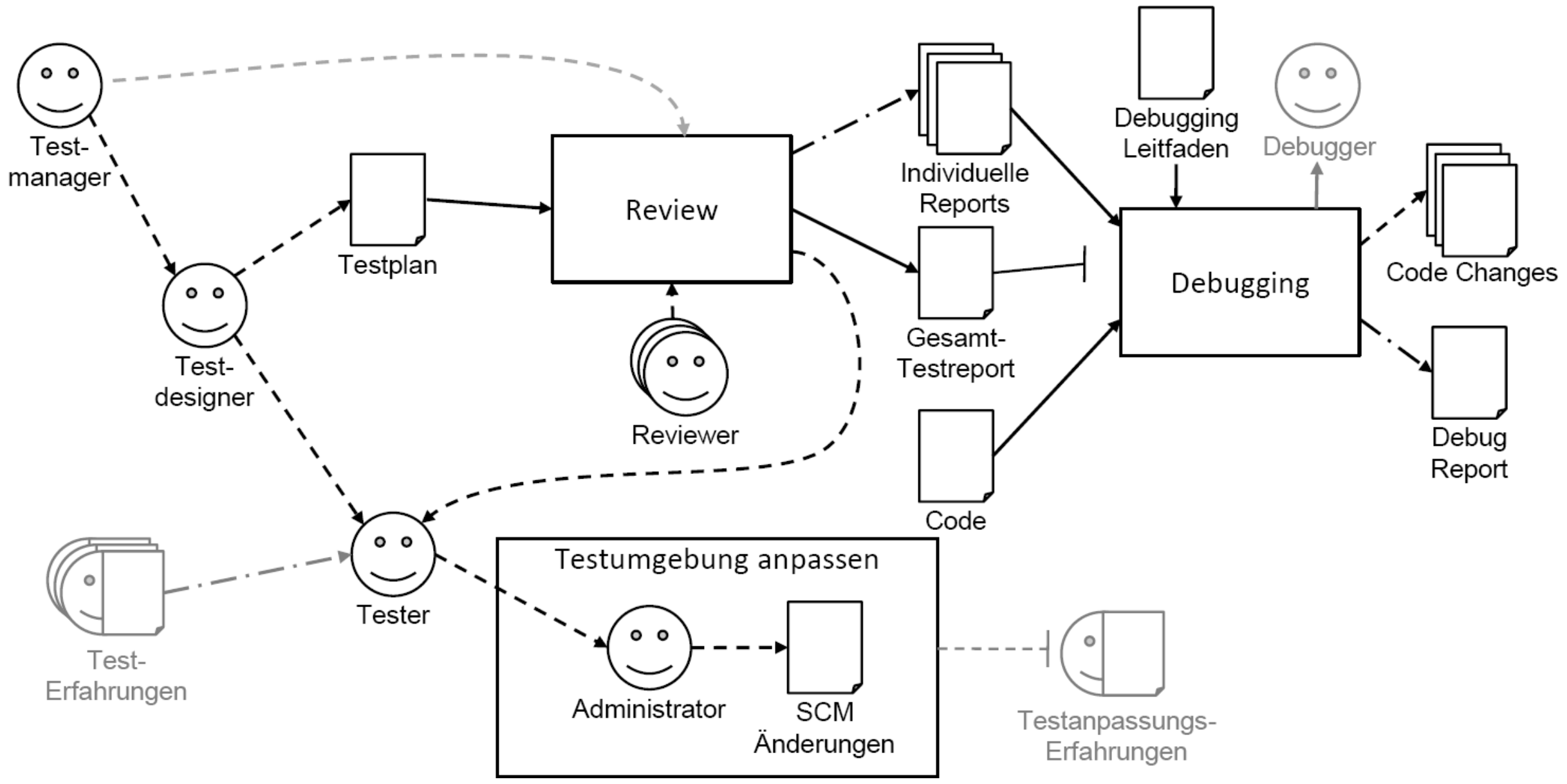

**Abbildung 1. Beispiel FLOW-Modell**

Es gibt die Möglichkeit Informationsflüsse von unterschiedlichen Seiten mit dem Aktivitätssymbol zu verbinden, um den Unterschied zwischen steuernden und inhaltlichen Informationsflüssen darstellen zu können. Inhaltliche Informationsflüsse enthalten die eigentlichen produktrelevanten Informationen wie Anforderungen, Entwurfsentscheidungen, Quellcode oder Testfälle, also meist Informationen darüber, wie das Problem aus der Anwendungsdomäne gelöst werden soll. Steuernde Informationsflüsse enthalten Informationen, die zur Durchführung der inhaltlichen Flüsse notwendig oder zumindest dafür hilfreich sind, also meist Informationen darüber, wie das Problem der Softwareentwicklung gelöst werden soll. Informationen, die steuernd an einer Aktivität mitwirken, werden von oben oder unten an das Aktivitätssymbol gezeichnet. Steuernde Informationen sind oft Erfahrungen. Informationen, die inhaltlich in der Aktivität verarbeitet werden, werden von links oder rechts an das Aktivitätssymbol gezeichnet.

Abbildung 1 zeigt ein Beispiel eines FLOW-Modells, in dem alle Modellelemente aus Tabelle 2 vorkommen.

## Modellierungsregeln

Folgende Regeln sollen die Erstellung von Informationsflussmodellen mit der FLOW-Notation unterstützen. Die Hinweise stammen aus unserer Erfahrung. Häufig werden bereits auf *syntaktischer* Ebene Fehler gemacht. In Tabelle 3 werden sowohl syntaktische als auch inhaltliche Modellierungsregeln mit je einem Beispiel dargestellt.

**Tabelle 3. Syntaktische und inhaltliche Modellierungsregeln von FLOW**

| Syntaktische Modellierungsregeln | Beispiel |
|---|---|
| Flüssige Speicher resultieren immer in flüssigen Flüssen (gestrichelter Pfeil). | 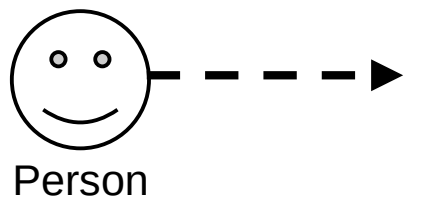  |
| Feste Speicher resultieren immer in festen Flüssen (durchgängiger Pfeil). | 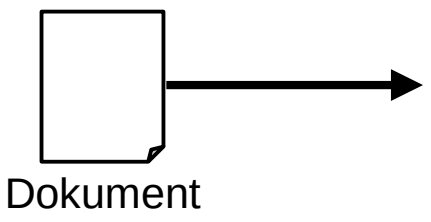  |
| Projektinformationen werden von links oder rechts an die Aktivität gezeichnet.<br>Steuernde Informationen (z.B. Erfahrungen) werden von oben oder unten an die Aktivität gezeichnet. | 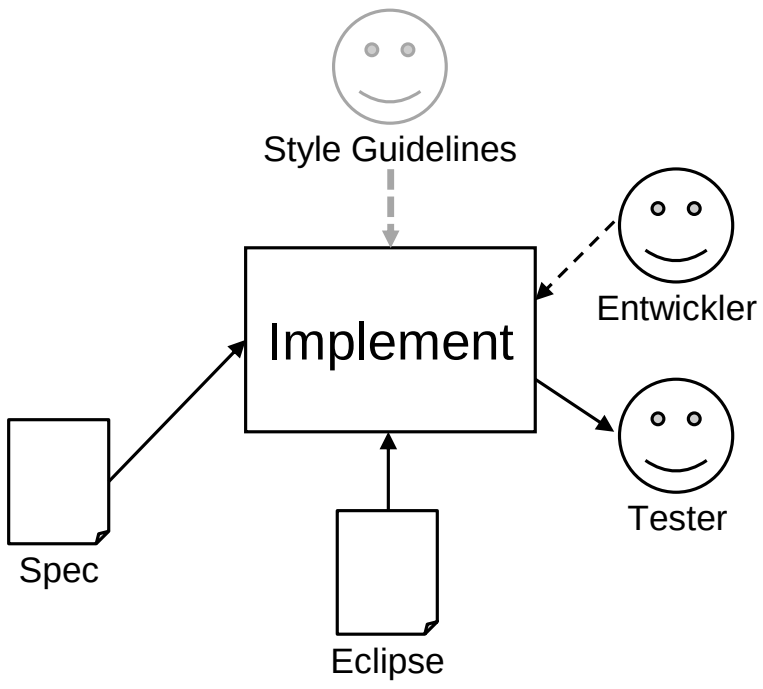  |

| Inhaltliche Modellierungsregeln | Beispiel |
|---|---|
| Zunächst nur grob modellieren und das Ziel im Auge behalten. Ziel ist meist *Verständnis* und nicht Vollständigkeit. Bei Bedarf kann das Modell später noch verfeinert werden. | |
| Wenn die genauen Flüsse innerhalb einer Aktivität nicht bekannt sind (Black-Box), werden ausgehende Flüsse per Default unbekannt, d.h. gestrichpunktet. Wenn der Typ eines ausgehenden Flusses bekannt ist, kann dieser auch explizit dargestellt werden. | |
| Die FLOW-Aktivität kann zur Strukturierung und Hierarchisierung von FLOW-Modellen genutzt werden, da sie Detailflüsse verbirgt. | |
| Üblicherweise werden entweder nur Rollennamen oder nur Personennamen als Bezeichner für flüssige Speicher in einem FLOW-Modell verwendet. Für den Fall, dass Beides in einem Modell vorkommt, kann man z. B. durch Unterstreichen die Rollen von den Individuen unterscheiden. | 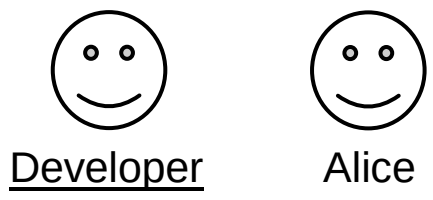  |
| FLOW-Modelle sind zeitlos, wie Datenfluss-Diagramme (DFDs). Der zu modellierende Zeitrahmen muss vorher festgelegt werden. | |
| Dennoch können zeitliche Abhängigkeiten aus der Reihenfolge von Flüssen abgeleitet werden, wenn keine Zyklen bestehen. | |

# 3 FLOW-Methode

Die FLOW-Methode leitet die *Analyse* und *Verbesserung* von Informationsflüssen in der Softwareentwicklung an, indem sie den prinzipiellen Ablauf eines FLOW-Verbesserungsvorhabens beschreibt. Zudem beschreibt die FLOW-Methode, wie bei einem FLOW-Verbesserungsvorhaben geeignete FLOW-Techniken für die Umsetzung gewählt werden.

Kern der Methode ist ein dreiphasiger Verbesserungsprozess (Abbildung 2), der Strategien und Verfahren zur (1) Erhebung, (2) Analyse und (3) Verbesserung von Informationsflüssen beschreibt. Im Forschungsprojekt InfoFLOW wurden verschiedene Techniken und Werkzeuge entwickelt (vgl. Kapitel 4), die Aktivitäten in den einzelnen Phasen des Verbesserungsprozesses beschreiben bzw. unterstützen. Um die richtigen Techniken und Werkzeuge auswählen zu können, müssen zunächst das verfolgte FLOW-Ziel und Randbedingungen klar sein. Die Festlegung des FLOW-Ziels und das Festhalten der Randbedingungen des Verbesserungsvorhabens geschieht in der Vorbereitungsphase (vgl. Kapitel 3.2).

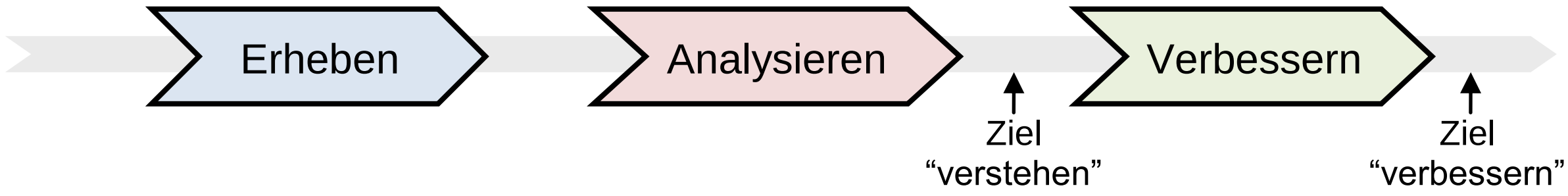

**Abbildung 2. Die 3 Phasen des FLOW-Verbesserungsprozesses.**

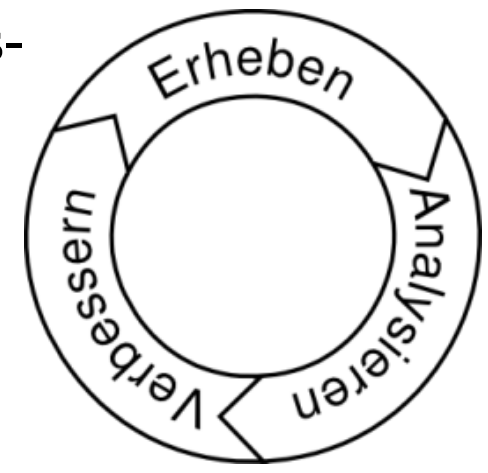

Die drei Phasen *Erheben*, *Analysieren*, und *Verbessern* leiten den Verbesserungsprozess an. Sie können iterativ (siehe rechts) ausgeführt werden, sodass zum Beispiel die Auswirkungen eines ersten Verbesserungsvorhabens in einem zweiten Durchlauf erhoben und analysiert werden können. Der FLOW-Verbesserungsprozess kann an verschiedenen Punkten begonnen und wieder verlassen werden. Zum Beispiel kann direkt mit der *Verbesserung* begonnen werden, wenn bereits eine *Analyse* ohne FLOW-Techniken stattgefunden hat.

FLOW bietet spezielle *Techniken* und *Werkzeuge* für jede der drei Phasen (siehe nächstes Kapitel). Um geeignete Techniken und Werkzeuge auswählen zu können sollte vor Einstieg in den FLOW-Verbesserungsprozess eine Vorbereitungsphase durchlaufen werden. In der Vorbereitung wird das zu Verfolgende FLOW-Ziel gewählt und Projektparameter wie Teamgröße und Zieldomäne erfasst.

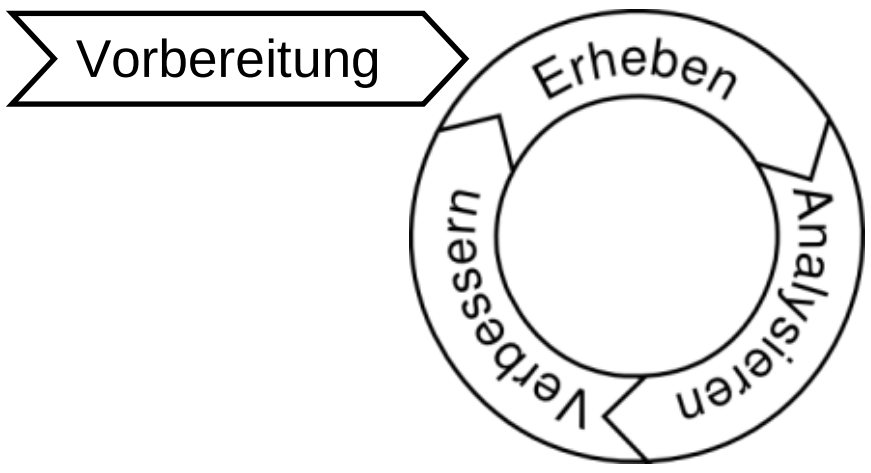

**Abbildung 3. Einordnung der Vorbereitungsphase in die FLOW-Methode**

## 3.1 Verbesserung der Softwareentwicklung mit der FLOW-Methode

An dieser Stelle wird kurz das grobe Vorgehen in einem Verbesserungsvorhaben unter Anwendung der FLOW-Methode skizziert. In den folgenden Kapiteln werden dann die Details der Phasen aus Abbildung 3 erläutert.

1) Verbesserungsvorhaben vorbereiten (vgl. Kapitel 3.2)
   a) FLOW-Ziel festlegen
   b) Projektparameter festhalten
   c) FLOW-Techniken auswählen (vgl. Kapitel 4)

2) Entsprechend der Beschreibungen der ausgewählten FLOW-Techniken den FLOW-Verbesserungsprozess (vgl. Abbildung 2) durchführen.
   a) Informationsflüsse erheben (vgl. Kapitel 3.3)
   b) Informationsflüsse analysieren (vgl. Kapitel 3.4)
   c) Informationsflüsse verbessern (vgl. Kapitel 3.5)
3) Ggf. weitere Iterationen des Verbesserungsprozesses durchführen, um die Effektivität von Maßnahmen vorheriger Iterationen zu evaluieren oder kontinuierliche Verbesserungen zu erreichen (vgl. Abbildung 3).

## 3.2 Phase 0: Vorbereitung

In der Vorbereitungsphase sind einige wichtige Grundlagen für den folgenden Verbesserungsprozess zu legen:

1. **Festlegung des im Verbesserungsprozess zu verfolgenden FLOW-Ziels.** Bei der Festlegung des FLOW-Ziels werden Absicht, Umfang und Zeit des Verbesserungsvorhabens festgelegt. Also z.B., ob verbessert oder nur analysiert werden soll, ob eine Aktivität oder ein ganzes Projekt betrachtet werden soll und ob der Verbesserungsprozess vor oder zur Projektlaufzeit durchgeführt werden soll.
2. **Projektparameter (bzw. Aktivitäts- oder Organisationsparameter) werden festgehalten.** Für die Auswahl von FLOW-Techniken relevante Projektparameter müssen festgehalten werden. Relevante Parameter sind z.B. die Anzahl der beteiligten Personen, die Rollenbesetzungen, das eingesetzte Vorgehensmodell und zu beachtende Randbedingungen (Einschränkungen, Verteiltheit, gesetzliche Vorgaben, vorhandene Ressourcen).
3. **Auswahl passender FLOW-Techniken:** Schließlich werden basierend auf dem FLOW-Ziel und den Projektparametern passende FLOW-Techniken ausgewählt.

Diese Vorbereitungsmaßnahmen sind notwendig, um später geeignete Techniken und Werkzeuge auswählen zu können. Im Folgenden wird genauer auf diese drei Schritte eingegangen.

### Schritt 1. FLOW-Ziel festlegen

Das Hauptziel der FLOW-Methode ist es *Informationsflüsse in der Softwareentwicklung zu verbessern*. Dieses Ziel kann erreicht werden, indem mindestens eines der Unterziele verfolgt wird, die aus den in Tabelle 4 dargestellten Aspekten eines FLOW-Ziels gebildet werden können.

**Tabelle 4. Drei Aspekte eines FLOW-Ziels.**

| | |
|---|---|
| **Absicht** | Verstehen, verbessern |
| **Zeit** | Vorher, während dessen, nachher |
| **Umfang** | Aktivität, Projekt, Organisation |

Die drei Aspekte eines FLOW-Ziels:

1. **Absicht:** Absicht eines FLOW-Ziels ist es entweder Informationsflüsse zu *verstehen* oder sie zu *verbessern*. Normalerweise erfüllt eine FLOW-Technik nacheinander beide Absichten: Zuerst müssen Informationsflüsse erhoben werden. Diese Informationsflüsse können dann analysiert und damit erst verstanden werden. Erst dann können die Informationsflüsse verbessert werden.
2. **Umfang:** Der Umfang eines FLOW-Ziels legt die Reichweite eines Verbesserungsvorhabens fest. Sie kann entweder eine *Aktivität*, ein *Projekt* oder eine ganze *Organisation* umfassen. Ist der Umfang auf eine Aktivität festgelegt, werden nur die Informationsflüsse einer einzelnen Softwareentwicklungsaktivität berücksichtigt, z.B. die Anforderungs-Elicitation. Auf Projektebene werden alle Informationsflüsse eines gesamten Softwareentwicklungsprojekts betrachtet. Wohingegen auf der Organisationsebene projektübergreifende Informationsflüsse im Mittelpunkt stehen.

3. **Zeit:** Die Hauptaktivitäten einer FLOW-Technik werden entweder *vor*, *nach* oder *während* der Durchführung der im Fokus stehenden Aktivität (bzw. des Projekts oder der Organisation) durchgeführt. Dieser zeitliche Aspekt ist Teil des FLOW-Ziels.

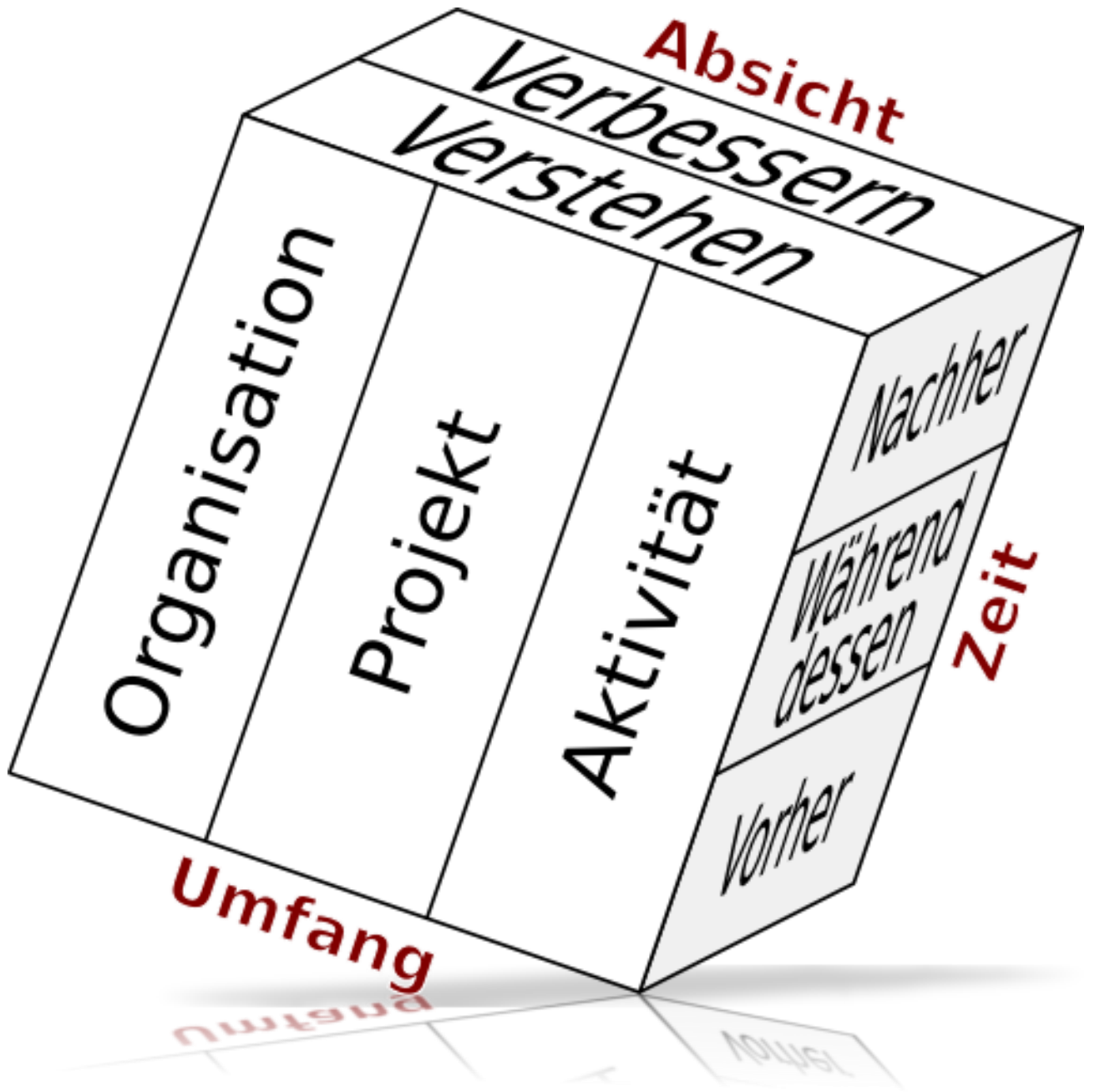

**Abbildung 4. FLOW-Goal-Cube: Die 3 Dimensionen eines FLOW-Ziels**

Abbildung 4 zeigt die drei Dimensionen eines FLOW-Ziels im sogenannten FLOW-Goal-Cube. Theoretisch gibt es 18 Permutationen dieser drei FLOW-Ziel-Aspekte. Einige Kombinationen sind aber nicht sinnvoll, z.B. Verbesserungsvorhaben, die nach der Lebensdauer einer Organisation durchgeführt werden sollen. Beispiele typischer FLOW-Ziele sind in Tabelle 5 aufgelistet.

**Tabelle 5. Beispiele von FLOW-Zielen.**

| FLOW-Ziel | Konkretes Ziel |
|---|---|
| Verstehe eine gerade durchgeführte Aktivität. | Verstehe die Anforderungserhebung während eines Interviews. |
| Verstehe eine Aktivität im Nachhinein. | Verstehe eine Reviewsitzung in einer Post-Mortem-Analyse. |
| Verbessere eine Aktivität im Voraus. | Spezifikationsreview vorher tailorn. |
| Verbessere ein gerade laufendes Projekt. | Verbessere Anforderungsfluss während der RE-Phase eines Projekts. |

## Schritt 2: Projektparameter sammeln

Zur Wahl geeigneter FLOW-Techniken ist neben dem verfolgten Ziel noch die Kenntnis einiger Projektparameter wichtig. Relevante Parameter sind:

1. **Projektgröße:** Die Projektgröße kann über die Teamgröße oder die Größe des verfügbaren Budgets angegeben werden. Beides ist wichtig, um abschätzen zu können, wie komplex die FLOW-Modelle wahrscheinlich werden und wie viele Ressourcen zur Anwendung der FLOW-Methode bereit stehen. Außerdem hat die Projektgröße Einfluss auf Kommunikations- und Dokumentationsaufwand. Daher können FLOW-Techniken für bestimmte Projektgrößen spezialisiert sein.
2. **Domäne:** FLOW-Techniken können auf bestimmte Anwendungsdomänen oder bestimmte Softwaredomänen (Web, Embedded, etc.) spezialisiert sein.
3. **Vorgehensmodell:** Die Art wie entwickelt wird, ob agil oder Prozess-getrieben, hat großen Einfluss auf Art und Häufigkeit von Informationsflüssen. Daher können FLOW-Techniken für bestimmte Vorgehensmodelle spezialisiert sein.

4. **Verteiltheit:** Informationsflüsse in lokalen Projekten sind meist anders als Informationsflüsse in verteilten Projekten. Einige FLOW-Techniken wurden speziell für verteilte Projekte entwickelt. Eine weitere wichtige Information für die Wahl einer geeigneten FLOW-Technik ist die Art der Verteilung. Informationsflüsse in Projekten, die entlang des Entwicklungsprozesses verteilt sind (z.B. Anforderungsphase in D, Implementierung in CZ), sind anders als Informationsflüsse in Projekten, die innerhalb einer Phase verteilt sind (Z.B. Implementierung auf drei Standorte verteilt), da die parallele Arbeit an einem Artefakt mehr koordinierende Kommunikation erfordert als die sequentielle.
5. **Sonstige Randbedingungen:** Andere Randbedingungen, die bei der Wahl einer geeigneten FLOW-Technik eine Rolle spielen können.

### Schritt 3: FLOW-Techniken auswählen

Der letzte Schritt der Vorbereitungsphase ist die Wahl der FLOW-Techniken, die zum gesetzten Ziel und den Projektparametern passen. Um dies zu vereinfachen sollten alle verfügbaren FLOW-Techniken mit Hilfe des in Tabelle 6 dargestellten Templates klassifiziert werden. Jede FLOW-Technik hat einen beschreibenden Namen. Eine FLOW-Technik bietet Unterstützung für verschiedene Aspekte eines FLOW-Ziels. Welche Aspekte das sind wird im Abschnitt „Ziel" des Templates aus Tabelle 6 markiert. Hier ist eine Mehrfachauswahl möglich, da es Techniken gibt, die zur Erreichung verschiedener FLOW-Ziel-Aspekte genutzt werden können, z.B. zur Verbesserung einer Entwicklungsaktivität oder zur Verbesserung eines ganzen Projekts. Im nächsten Absatz des Templates werden die Phasen des FLOW-Verbesserungsprozesses notiert, für die die FLOW-Technik konkrete Unterstützung anbietet. Der letzte Teil des Templates sind die Projektparameter, für die eine Technik entwickelt wurde bzw. die den Anwendungsbereich einer Technik einschränken. Für ein gegebenes FLOW-Ziel und gegebene Randbedingungen können dann geeignete Techniken durch Abgleich der klassifizierten FLOW-Techniken mit dem Ergebnis der Vorbereitungsphase gefunden werden. Im Folgenden werden die wesentlichen Teilschritte der Wahl geeigneter FLOW-Techniken noch einmal zusammen gefasst:

1. **Vorbedingung der FLOW-Technik-Wahl:** Alle verfügbaren FLOW-Techniken (vgl. Kapitel 4) sind mit dem Template aus Tabelle 6 klassifiziert.
2. **Vorbereitungsergebnis dokumentieren:** Für das aktuelle Verbesserungsvorhaben das Template aus Tabelle 6 ausfüllen. Dabei werden die Ergebnisse aus den beiden vorangegangenen Schritten (FLOW-Ziel festlegen und Projektparameter sammeln) dokumentiert.
3. **FLOW-Techniken wählen:** Basierend auf den ausgefüllten Templates (aktuelles Vorhaben vs. Vorhandene Techniken) wird ein Abgleich des geplanten Verbesserungsvorhabens mit den zur Verfügung stehenden FLOW-Techniken durchgeführt:
   - Es kommen nur die FLOW-Techniken in Frage, die zum gesteckten **FLOW-Ziel** passen (Absicht, Zeit, Umfang).
   - Weiterhin müssen die **Projektparameter** vom geplanten Verbesserungsvorhaben zu den FLOW-Techniken passen, d.h. alle Techniken, die nicht dazu passen werden nicht weiter betrachtet. Z.B. können Techniken, die nur für sehr kleine Projekte geeignet sind (weniger als 8 Entwickler), nicht zur Verbesserung eines 100-Personen-Projekts eingesetzt werden.
   - Schließlich müssen aus den verbleibenden FLOW-Techniken diejenigen gewählt werden, die Aktivitäten in den **Phasen** unterstützen, die zur Erreichung des gesteckten FLOW-Ziels durchlaufen werden sollten. Ist die Absicht eine Verbesserung, so sollten alle 3 Phasen durchlaufen werden. Ist die Absicht Verstehen, so reicht es, wenn die ersten beiden Phasen durchlaufen werden. Um eine Abdeckung aller notwendigen Phasen zu erreichen, können FLOW-Techniken auch kombiniert werden, z.B. die Informationsfluss-Elicitation (vgl. Kapitel 4.1) für die Erhebungsphase und die Mustersuche (vgl. Kapitel 4.7) für die Analysephase.

Wenn alle Vorbedingungen (vgl. Kapitel 1.2) erfüllt sind und nachdem die Vorbereitungsphase durchgeführt wurde, kann der FLOW-Verbesserungsprozess (vgl. Abbildung 3) gestartet werden.

**Tabelle 6. Template zur Vorbereitung des FLOW-Verbesserungsprozesses**

**Vorbereitung des FLOW-Verbesserungsprozesses**

**Name des Vorhabens:** ________________________ **Verantwortliche(r):** __________________________

| | | | | |
|---|---|---|---|---|
| **Ziel** 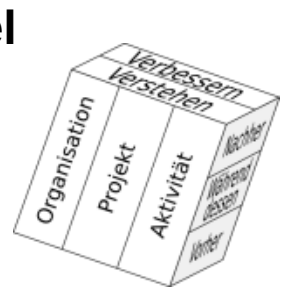  | **Absicht** | ☐ Verstehen | ☐ Verbessern | |
| | **Zeit** | ☐ Vorher | ☐ Während dessen | ☐ Nachher |
| | **Umfang** | ☐ Aktivität<br>______________ | ☐ Projekt<br>______________ | ☐ Organisation<br>______________ |
| **Phasen** 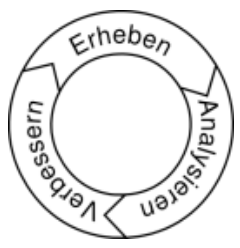  | | ☐ Erheben<br>☐ Analysieren<br>☐ Verbessern | | |
| **Projektparameter** | **Projektgröße** | ☐ Teamgröße: ____________ | ☐ Budget: __________________ | |
| | **Domäne** | _______________________________________________________ | | |
| | **Vorgehensmodell** | ☐ Agil: ________________ | ☐ Prozess: __________________ | |
| | **Verteiltheit** | ☐ lokal ☐ verteilt (☐ vertikal ☐ horizontal ☐ sonst. _________) | | |
| | **Sonstiges** | _______________________________________________________<br>_______________________________________________________<br>_______________________________________________________ | | |

## 3.3 Phase 1: Erheben

Die erste Phase des FLOW-Verbesserungsprozesses hat die Erhebung von Informationsflüssen zum Ziel. Es gilt eine Basis für spätere Analysen und Verbesserungen zu schaffen. Ergebnis der ersten Phase ist ein FLOW-Modell, das in den folgenden Phasen zur Analyse und als Ausgangspunkt für eine Verbesserung genutzt werden kann. Als Erheben gilt sowohl das Erstellen eines FLOW-Modells ausgehend von vorhandenen Informationsflüssen (Ist-Modell), zum Beispiel durch Beobachtung, als auch das konstruktive planende Erstellen eines FLOW-Modells (Soll-Modell).

Techniken und Werkzeuge der 1. Phase lassen sich nach zwei Aspekten klassifizieren:

1. nach der *Strategie* und
2. nach dem *Verfahren* zur Datenerfassung.

### 3.3.1. Erhebungsstrategien

Die Strategie bestimmt die Herangehensweise an die Erhebung der Informationsflüsse. Es gibt die *Top-Down-* und die *Bottom-Up-*Strategie. Bei der Top-Down-Strategie werden ausgehend von einem Informationsflussmodell auf hoher Abstraktionsebene Schritt für Schritt Informationsflussdetails hinzugefügt. Dies führt zu einem immer ausführlicheren Modell auf niedrigerer Abstraktionsebene. Bei der Bottom-Up-Strategie wird durch schrittweise Zusammenführung mehrerer unabhängiger Einzelmodelle auf niedriger Abstraktionsebene nach und nach ein Gesamtmodell höherer Abstraktion erstellt.

Die Entscheidung, welche Strategie gewählt werden sollte, wird vom Aspekt „Umfang“ des FLOW-Ziels beeinflusst. So ist es zum Beispiel sinnvoll bei einer organisationsweiten Betrachtung der Informationsflüsse den Top-Down-Ansatz zu wählen und ausgehend von einem abstrakten Überblicksdia-gramm, das die Informationsflüsse der gesamten Organisation zeigt, an interessanten Stellen Details hinzuzufügen, um diese genauer Analysieren zu können. Eine weitere Entscheidungshilfe zur Wahl einer geeigneten Erhebungsstrategie bieten die Vor- und Nachteile in Tabelle 7.

**Tabelle 7. Vor- und Nachteile der Strategien der Phase 1: Erheben**

| | Vorteile | Nachteile |
|---|---|---|
| **Bottom-Up** | • Zu Beginn kein Gesamtüberblick notwendig<br>• Eignet sich gut für Ist-Modelle | • Zusammenführen der Einzelmodell oft nicht trivial<br>• Lokale Sichten können stark differieren |
| **Top-Down** | • Bietet Möglichkeit nur bei relevanten Stellen ins Detail zu gehen<br>• Eignet sich gut für Soll-Modelle | • Gesamtüberblick ist häufig nicht vorhanden<br>• Zu große Modelle können unübersichtlich sein |

## 3.3.2. Erhebungsverfahren

Verfahren für die Erhebung von Informationsflüssen beschreiben typische Vorgehensweisen, wie man die Daten, die zur Erstellung eines Informationsflussmodells notwendig sind, sammelt und umwandelt. Erhebungsverfahren lassen sich nach der Datenquelle unterscheiden (vgl. Abbildung 5): Elicitation-Verfahren nutzen das Wissen von Personen über die zu erhebenden Informationsflüsse. Ableitende Verfahren nutzen bereits bestehende Modelle, die Aspekte von Informationsflüssen enthalten, z.B. Prozessmodelle. Eine FLOW-Technik kann auch mehrere Erhebungsverfahren kombinieren, z.B. eine Modellableitung und eine anschließende Elicitation.

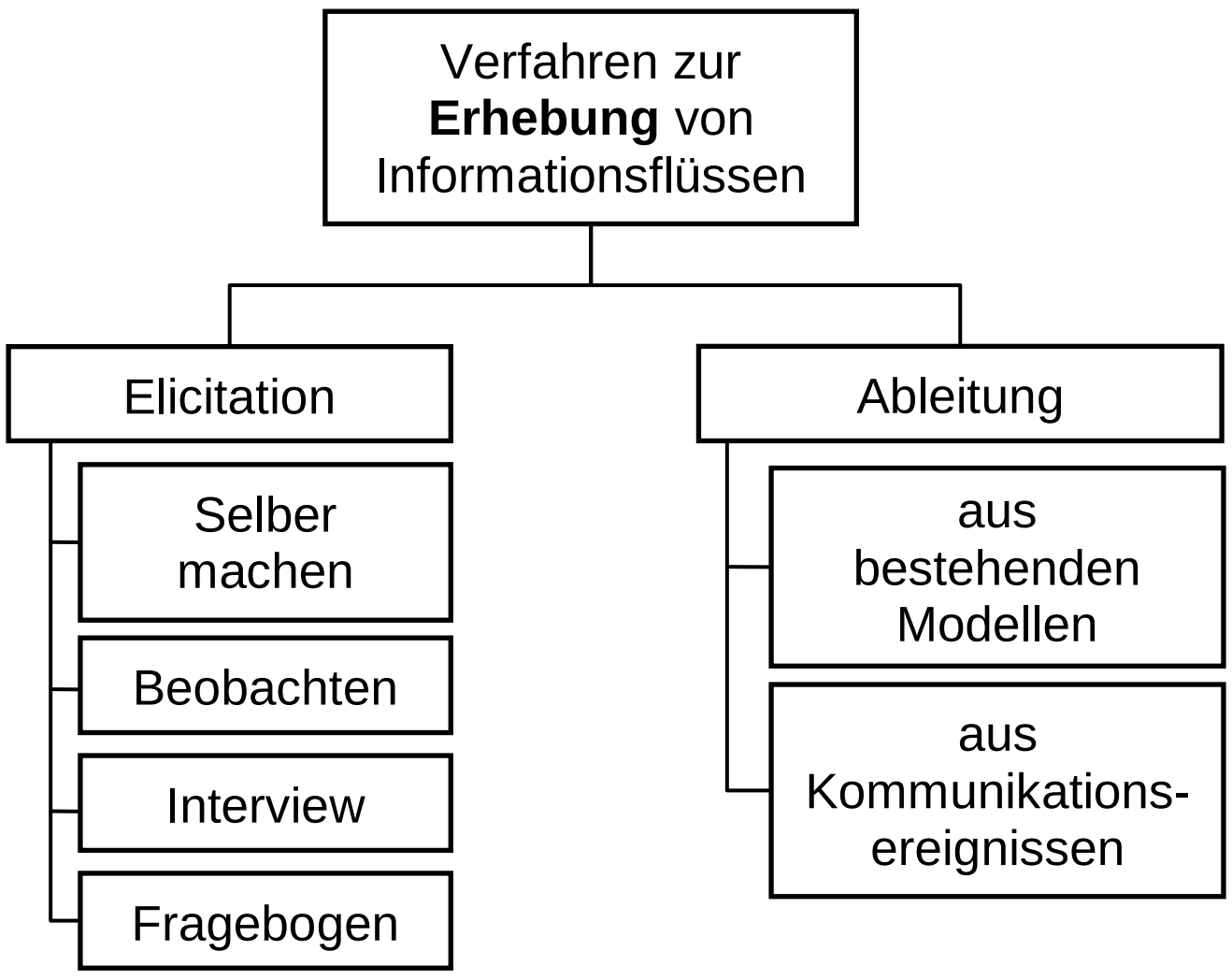

**Abbildung 5. Verfahren zur Erhebung von Informationsflüssen**

Verfahren der Elicitation sind (vgl. Abbildung 5):

- **Selber machen:** Ein FLOW-Experte begibt sich selbst in die Situation, zu der Informationsflüsse erhoben werden sollen, und führt diese selbst aus. Dabei notiert er alle relevanten Informationsflüsse und erstellt das Informationsflussmodell.
- **Beobachten:** Ein FLOW-Experte beobachtet die Situation, zu der Informationsflüsse erhoben werden sollen. Dabei notiert er alle relevanten Informationsflüsse und erstellt das Informationsflussmodell.
- **Interview:** Ein FLOW-Experte führt mit allen – oder zumindest den wichtigsten – Personen ein Interview, die an der Situation beteiligt sind, zu der Informationsflüsse erhoben werden sollen. Dabei notiert er alle relevanten Informationsflüsse und erstellt das Informationsflussmodell.
- **Fragebogen:** Ein FLOW-Experte erstellt einen Fragebogen, der alle Informationsfluss-relevanten Fragen enthält, und verteilt diese an möglichst alle Personen, die an der Situation beteiligt sind, zu der Informationsflüsse erhoben werden sollen. Dabei notiert er alle relevanten Informationsflüsse und erstellt das Informationsflussmodell.

Verfahren der Ableitung sind:

- **Modellableitung:** Ausgehend von einem bestehenden Modell, das Informationsflussaspekte der zu untersuchenden Situation enthält (z.B. ein Prozessmodell), werden relevante Teile extrahiert und in einem neuen Informationsflussmodell zusammen gefasst.
- **Kommunikationsereignisableitung:** Aus Kommunikationsereignissen wie E-Mails, Telefonaten, Meetings etc. werden Informationsflüsse abgeleitet.

Bei allen Erhebungsverfahren ist zu beachten, dass die gewonnenen Daten nur indirekt Aufschluss über tatsächliche Informationsflüsse geben. Z.B. können Interviewdaten subjektiv verfälscht und modellierte Prozessmodelle veraltet sein. Kommunikationsereignisse sagen nicht direkt etwas über den kommunizierten Inhalt aus. Aus einem Kommunikationsereignis, wie zum Beispiel einer Skype-Telefonkonferenz, das aus Start-, Endzeitpunkt und einer Liste von Teilnehmern besteht, kann nicht direkt ein Informationsfluss abgeleitet werden. Es kann entweder gar keine Information fließen, zum Beispiel wenn jeder Teilnehmer eine andere Sprache spricht, oder es fließt nicht projektrelevante Information, zum Beispiel wenn über das letzte Freizeiterlebnis gesprochen wird. Eine Ableitung von Informationsflüssen aus Kommunikationsereignissen funktioniert also nur unter gewissen Annahmen:

- Es wird angenommen, dass während einer Kommunikation auch Informationen fließen.
- Es wird angenommen, dass während einer professionellen Kommunikation hauptsächlich über projektrelevante Dinge gesprochen wird. Davon ist im Softwareentwicklungsumfeld auszugehen. So konnte in einer Studie über das Kommunikationsverhalten beim Pair-Programming gezeigt werden, dass nur 7% der Kommunikation privaten Inhalt haben [19].

Die folgende Tabelle 8 listet einige Vor- und Nachteile der einzelnen Verfahren auf. Diese gilt es bei der Auswahl des anzuwendenden Verfahrens zu berücksichtigen.

**Tabelle 8. Vor- und Nachteile der Verfahren der Phase 1: Erheben**

| | Vorteile | Nachteile |
|---|---|---|
| **Selber machen** | • Sehr effektiv. Kein Stille-Post-Effekt verfälscht Ergebnis<br>• Ermöglicht auch implizites Wissen zu erfassen | • Teuer<br>• Oft nicht möglich, da<br>○ Domänenwissen fehlt<br>○ Nicht erlaubt |
| **Beobachtung** | • Effektiv. Kein Stille-Post-Effekt verfälscht Ergebnis | • Teuer<br>• Oft nicht möglich, da nicht erlaubt |
| **Interview** | • Flexibel. Es kann auf vorher unbekannte Sachverhalte eingegangen werden | • Hoher Vorbereitungsaufwand |
| **Fragebogen** | • Günstig umzusetzen<br>• Erleichtert Auswertung | • Hoher Vorbereitungsaufwand<br>• Starr, nicht möglich auf vorher nicht bekannte Sachverhalte einzugehen |
| **Modellableitung** | • Wiederverwendung bestehender Modelle<br>• Vergleichsweise günstig<br>• Kann teilweise automatisiert werden | • Deckt meist nur Teilaspekte von Informationsflüssen ab<br>• Kann je nach Ausgangsmodell große unübersichtliche Informationsflussmodelle zur Folge haben |
| **Kommunikations-ereignisableitung** | • Kann teilweise automatisiert werden<br>• Nutzung bestehender Kommunikationsinfrastruktur möglich<br>• Bei Automatisierung vergleichsweise günstig | • Es ist schwierig tatsächliche Informationsflüsse aus Ereignissen herzuleiten<br>• Automatische Ableitung kann viel Vorbereitungsaufwand erfordern<br>• Automatische Erfassung kann zu Datenschutzproblemen führen |

Tabelle 9 fasst Strategien und Verfahren der ersten Phase des FLOW-Verbesserungsprozesses zusammen.

**Tabelle 9. Zusammenfassung der phasenspezifischen Aspekte für die Phase 1: Erheben**

| **Phasenspezifische Aspekte Phase 1: Erheben** | | | | |
|---|---|---|---|---|
| **Strategie** | ☐ Bottom-Up | ☐ Top-Down | | |
| **Verfahren** | ☐ Selber machen<br>☐ Modellableitung | ☐ Beobachten<br>☐ Kommunikationsereignisableitung | ☐ Interview | ☐ Fragebogen |

## 3.4 Phase 2: Analysieren

Bei der Analyse geht es darum die Informationsflüsse zu verstehen und Probleme zu finden. Das wichtigste Werkzeug der Analyse ist die Visualisierung der Informationsflüsse. Prinzipiell ist zwischen manueller und automatischer Analyse zu unterscheiden. Die manuelle Analyse erfolgt durch Auswertung der Informationsflussmodelle durch einen FLOW-Experten. Eine automatische Analyse kann zum Beispiel mit Hilfe einer automatischen Pattern-Suche erfolgen. Dazu muss jedoch das vorliegende FLOW-Modell eine gewisse Reife haben und detailliert genug sein.

### 3.4.1. Analysestrategien

Die Analysestrategie bestimmt die prinzipielle Herangehensweise an die Analyse von Informationsflüssen. Es gibt die Strategie der manuellen, automatischen und teilautomatischen Analyse von Informationsflüssen. Bei der manuellen Strategie analysiert ein FLOW-Experte das Informationsflussmodell. Bei der (teil-)automatischen Analyse werden Probleme in Informationsflussmodellen (teil-) automatisch gesucht. Tabelle 10 zeigt die Vor- und Nachteile der beiden Pole der manuellen und automatischen Analysestrategien. Bei der teilautomatischen Analyse vermischen sich Vor- und Nachteile der beiden Strategien.

**Tabelle 10. Vor- und Nachteile der Strategien der Phase 2: Analysieren**

| | **Vorteile** | **Nachteile** |
|---|---|---|
| **manuell** | • Expertenwissen leichter nutzbar<br>• Weniger Vorbereitungsaufwand<br>• Zusatzwissen, welches nicht im Modell abgebildet ist, kann zur besseren Analyse mit genutzt werden | • Ergebnisse abhängig von FLOW-Experten<br>• Ergebnisse können subjektiv beeinflusst sein<br>• Nur für relativ kleine Modelle geeignet |
| **automatisch** | • Objektive Analysen möglich<br>• Ergebnisse leicht reproduzierbar<br>• Auch für große Modelle geeignet | • FLOW-Modelle müssen vollständig und in gewisser Form vorliegen<br>• Hoher Vorbereitungsaufwand zur Erstellung von Analysewerkzeugen |

### 3.4.2. Analyseverfahren

Verfahren zur Analyse von Informationsflüssen beschreiben typische Vorgehensweisen, wie Probleme und andere Besonderheiten in Informationsflussmodellen identifiziert werden können. Für die Analyse von Informationsflussmodellen stehen die folgenden drei Verfahren zur Verfügung (Abbildung 6):

- **Visualisierung:** Die Visualisierung ist das grundlegendste Analyseverfahren. Dabei wird ein FLOW-Modell mit Hilfe der FLOW-Notation visuell dargestellt und meist manuell (siehe Strategie) untersucht. Auffälligkeiten wie überwiegend feste oder flüssige Flüsse oder unnötige Dokumentation können so identifiziert werden (vgl. auch Mustersuche unten).
- **Simulation:** Bei der Simulation werden mit Hilfe von zusätzlichen Elementen im Informationsflussmodell, z.B. Informationseinheiten (vgl. u.a. [9]), Informationsflüsse simuliert. Mit Hilfe von Simulationen können dynamische Probleme, z.B. das Vergessen wichtiger Informationen, leichter erkannt werden.
- **Mustersuche:** Bei der Mustersuche werden FLOW-Muster (vgl. [5]) im FLOW-Modell gesucht. Je nach Strategie kann dies manuell, z.B. mit Hilfe des FLOW-Musterkatalogs [6], oder automatisch, z.B. mit Hilfe von ProFLOW [7], geschehen.

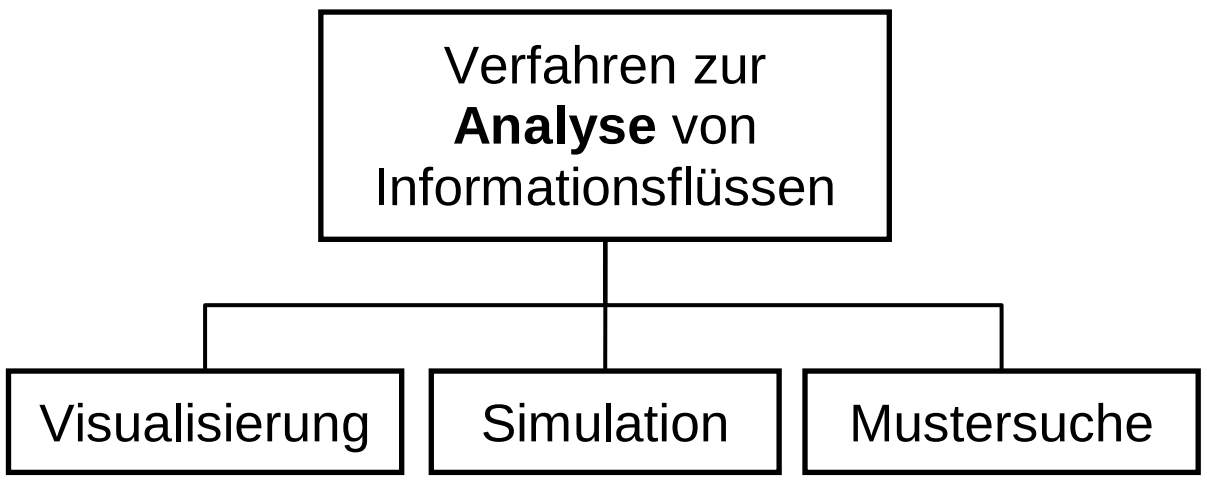

**Abbildung 6. Verfahren zur Analyse von Informationsflüssen**

Visualisierung, Simulation und Mustersuche können in einer FLOW-Technik auch kombiniert werden.

Tabelle 11 zeigt einige Vor- und Nachteile der unterschiedlichen Analyseverfahren.

**Tabelle 11. Vor- und Nachteile der Verfahren der Phase 2: Analysieren**

| | Vorteile | Nachteile |
|---|---|---|
| **Visualisierung** | • Erfahrung des Analysten wird genutzt | • Teurer Personalaufwand (manuell)<br>• Ineffektiv bei großen Modellen |
| **Simulation** | • Erfahrung des Analysten wird genutzt | • Teurer Personalaufwand (manuell)<br>• Simulation muss erstellt, vorbereitet und durchgeführt werden |
| **Mustersuche** | • Günstig und effektiv durch Automatisierung | • Abhängig von Quantität und Qualität der Muster<br>• Abhängig von Matching-Algorithmus |

Tabelle 12 fasst Strategien und Verfahren der zweiten Phase des FLOW-Verbesserungsprozesses zusammen.

**Tabelle 12. Zusammenfassung der phasenspezifischen Aspekte für die Phase 2: Analysieren**

| **Phasenspezifische Aspekte Phase 2: Analysieren** | | | |
|---|---|---|---|
| **Strategien** | ☐ manuelle Analyse | ☐ teilautomatische Analyse | ☐ automatische Analyse |
| **Verfahren** | ☐ Visualisierung | ☐ Simulation | ☐ Mustersuche |

## 3.5 Phase 3: Verbessern

Die letzte Phase im FLOW-Verbesserungsprozess ist die Verbesserung der Informationsflüsse auf Basis der Erkenntnisse der beiden vorangegangenen Phasen. Als Verbesserung im Sinne der dritten Phase gilt sowohl ein Soll-FLOW-Modell, welches die verbesserten aber noch nicht umgesetzten Informationsflüsse enthält, als auch tatsächlich verbesserte Informationsflüsse in realen Projekten, unabhängig davon ob diese modelliert sind oder nicht. Techniken der dritten Phase lassen sich nach Verbesserungsstrategie und Verbesserungsverfahren unterscheiden.

### 3.5.1. Verbesserungsstrategien

Die Strategie zur Verbesserung von Informationsflüssen bestimmt den prinzipiellen Ansatz, der von einer Technik verfolgt wird, um Informationsflüsse zu verbessern. Es gibt vier Verbesserungsstrategien, von denen sich jeweils zwei gegenseitig ausschließen: Hauptprodukt- und Nebenprodukt-Strategie, sowie schwergewichtige und leichtgewichtige Strategie. Bei der Hauptprodukt-Strategie wird eine Verbesserung durch zusätzlich neu zu erstellende Informationen angestrebt. Mit der Nebenprodukt-Strategie (vgl. [12]) wird versucht, ausgehend von vorhandenen Informationsflüssen, eine Verbesserung durch Ableitung von Informationsflüssen als Nebenprodukt zu erreichen, d.h. vorhandene Informationen wiederzuverwenden. Eine weitere Strategie ist die leichtgewichtige Strategie. Bei der leichtgewichtigen Strategie sollen Verbesserungen im Informationsfluss durch einen möglichst geringen Zusatzaufwand erreicht werden. Ihr Gegenteil ist die schwergewichtige Strategie. Dabei werden Verbesserungen mit hohem Aufwand angestrebt. Meist werden Nebenprodukt und leichtgewichtige Strategie zusammen verwendet, da Änderungen im Prozess oft nur dann durchsetzbar sind, wenn für diejenigen, die den Prozess ausführen, möglichst wenig Zusatzaufwand zur eigentlichen Arbeit entsteht. Tabelle 13 zeigt einige Vor- und Nachteile der beiden Verbesserungsstrategien.

**Tabelle 13. Vor- und Nachteile der Strategien der Phase 3: Verbessern**

| | Vorteile | Nachteile |
|---|---|---|
| **Nebenprodukt** | • Wiederverwendung vorhandener Informationen | • Einmalige Vorbereitungsaufwände (z.B. für Nebenprodukt-Werkzeuge) können teils sehr hoch sein |
| **Hauptprodukt** | • Neue bzw. andere Informationen können berücksichtigt werden | • Zusätzliche Informationen führen meist zu mehr Arbeitsaufwand |
| **Leichtgewichtig** | • Wenig zusätzlicher Arbeitsaufwand<br>• Leicht zu lernen<br>• Wenig Anpassungen notwendig<br>• Einfach durchzuführen | • Effekte könnten zu gering sein |
| **Schwergewichtig** | • Es können große Verbesserungseffekte erzielt werden | • Schwergewichtige Ansätze werden oft nicht angenommen, da hoher Zusatzaufwand abschreckt<br>• Hohe Einführungshürden bei Management und Entwicklern |

### 3.5.2. Verbesserungsverfahren

Verfahren zur Verbesserung von Informationsflüssen beschreiben typische Flussveränderungen, die einen übergeordneten Informationsfluss verbessern können. Prinzipiell lassen sich Verfahren zur Verbesserung in vier Kategorien aufteilen, je nach FLOW-Modellelement Speicher, Fluss und Aktivität (vgl. Abbildung 1 und Tabelle 2), oder Informationsflussmuster (vgl. Abbildung 7).

1. Aggregatzustandsänderung
    a. **Verfestigung:** Bei der Verfestigung wird eine Verbesserung angestrebt, indem flüssige Informationen in eine feste Form gebracht werden. Dabei sollen die Vorteile des festen Aggregatzustands, wie wiederholter und langfristiger Abruf oder Verständlichkeit für Dritte, ausgenutzt werden.

b. **Verflüssigung:** Bei der Verflüssigung wird eine Verbesserung angestrebt, indem feste Informationen in eine flüssige Form gebracht werden. Dabei sollen die Vorteile des flüssigen Aggregatzustands, wie effiziente und schnelle Weitergabe, ausgenutzt werden.

2. Flussveränderung

   a. **Abkürzung:** Bei der Abkürzung wird eine Verbesserung angestrebt, indem Informationsflüsse über weniger Zwischenschritte, d.h. weniger Zwischenspeicher, ans Ziel geleitet werden. Dabei soll eine schnellere Informationsweitergabe erreicht und Fehler vermieden werden.

   b. **Umweg:** Bei einem Umweg wird eine Verbesserung angestrebt, indem Informationsflüsse über mehr Zwischenschritte, d.h. mehr Zwischenspeicher, ans Ziel geleitet werden. Dabei soll eine breitere Informationsverteilung und zusätzlicher Input (z.B. Reviews) erreicht werden.

   c. **Verzweigung:** Bei der Verzweigung wird eine Verbesserung angestrebt, indem Informationsflüsse an mehr Informationsspeicher verteilt werden. Dabei soll eine breitere Informationsverteilung erreicht werden.

   d. **Zusammenfassung:** Bei der Zusammenfassung von Informationsflüssen wird eine Verbesserung angestrebt, indem Informationsflüsse an weniger Informationsspeicher verteilt werden. Dabei soll eine gezieltere und damit Ressourcen-schonendere Informationsverteilung erreicht werden.

3. Aktivitätsänderung

   a. **Schnittstellenanpassung:** Bei der Schnittstellenanpassung wird eine Verbesserung angestrebt, indem Aggregatzustand und Anzahl eingehender, ausgehender, steuernder und unterstützender Informationsflüsse angepasst werden. Dabei soll eine gezieltere, d.h. der Aufgabe angemessene, Informationsbereitstellung bzw. -erstellung erreicht werden. Eine geänderte Schnittstelle zeiht meist auch eine Änderung der Aktivität selbst nach sich (siehe 3.b.).

   b. **Aktivitätsanpassung:** Bei der Aktivitätsanpassung wird eine Verbesserung durch Veränderung der Interna einer Aktivität angestrebt. Die Schnittstelle nach außen bleibt gleich. Es werden nur interne Informationsflüsse verändert.

4. **Musterersetzung:** Bei der Musterersetzung werden Teilinformationsflüsse gezielt durch FLOW-Muster ersetzt, die gewünschte Informationsflusseigenschaften aufweisen (vgl. auch Mustersuche bei den Analyseverfahren).

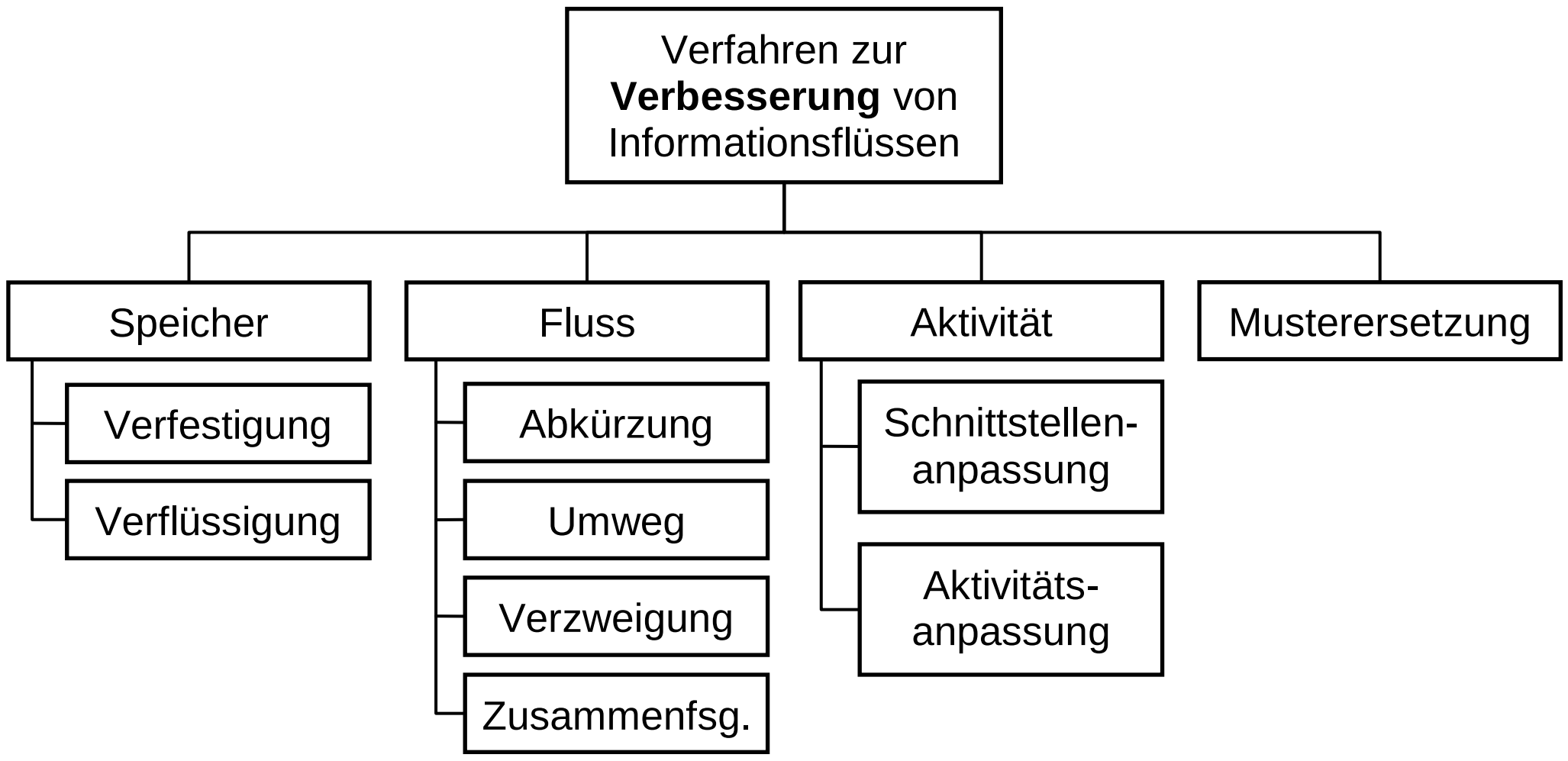

**Abbildung 7. Verfahren zur Verbesserung von Informationsflüssen**

Eine Verbesserungstechnik kann mehrere Verfahren kombinieren und so z.B. eine Verbesserung durch Verflüssigung und Abkürzung erzielen.

Die folgende Tabelle 14 listet einige Vor- und Nachteile der einzelnen Verfahren auf. Diese gilt es bei der Auswahl des anzuwendenden Verfahrens zu berücksichtigen.

**Tabelle 14. Vor- und Nachteile der Verfahren der Phase 3: Verbessern**

| | Vorteile | Nachteile |
|---|---|---|
| **Verfestigung** | • Wiederholter Zugriff möglich<br>• Langfristiger Zugriff möglich<br>• Verständlichkeit für Dritte | • Erstellung kostet Zeit und Aufwand<br>• Abruf kostet Zeit und Aufwand |
| **Verflüssigung** | • Effiziente und schnelle Informationsweitergabe | • Gehen leicht verloren<br>• Oft nicht für Dritte verständlich |
| **Abkürzung** | • Schnelle Informationsweitergabe<br>• Fehlervermeidung durch Verfälschung | • Weniger Informationsverteilung |
| **Umweg** | • Redundanz (mehr Sicherheit)<br>• Breitere Informationsverteilung<br>• Zusätzlicher Input / Validierungsmöglichkeit | • Redundanz (mehr Wartungsaufwand)<br>• Langsame Informationsweitergabe<br>• Fehler durch Verfälschung |
| **Verzweigung** | • Redundanz (mehr Sicherheit)<br>• Breitere Informationsverteilung<br>• Zusätzlicher Input / Validierungsmöglichkeit | • Redundanz (mehr Wartungsaufwand)<br>• Inkonsistenzen und Fehler durch Verfälschung |
| **Zusammenfassung** | • Redundanz (weniger Wartungsaufwand)<br>• Gezieltere Informationsverteilung | • Redundanz (weniger Sicherheit)<br>• Fehlende Validierungsmöglichkeiten |
| **Schnittstellenanpassung** | • Individuelle Informationsbereitstellung | • Hoher Vorbereitungsaufwand |

Tabelle 15 Fasst Strategien und Verfahren der dritten Phase des FLOW-Verbesserungsprozesses zusammen.

**Tabelle 15. Zusammenfassung der phasenspezifischen Aspekte für die Phase 3: Verbessern**

| **Phasenspezifische Aspekte Phase 3: Verbessern** | | | |
|---|---|---|---|
| **Strategie** | ○ Nebenprodukt<br>○ Leichtgewichtig | ○ Hauptprodukt<br>○ Schwergewichtig | |
| **Verfahren** | ☐ Aggregatzustand | ( ○ verfestigen | ○ verflüssigen ) |
| | ☐ Fluss | ( ○ Abkürzung<br>○ Verzweigung | ○ Umweg<br>○ Zusammenführung ) |
| | ☐ Aktivität | ( ○ Schnittstellen- | ○ Aktivitätsanpassung ) |
| | ☐ Musterersetzung | | |

# 4 FLOW-Techniken

Zur praktischen Durchführung und Unterstützung des FLOW-Verbesserungsprozesses gibt es verschiedene Techniken und zugehörige Werkzeuge. Techniken beschreiben für mindestens eine der drei Phasen des Verbesserungsprozesses ein Vorgehen (vgl. Abbildung 2). Dabei werden meist Strategien und Verfahren zu einer konkret anwendbaren Technik kombiniert. Die Beschreibungen der FLOW-Techniken enthalten Aktivitätsbeschreibungen zu jeder der unterstützten Phasen. Viele der Aktivitäten werden durch Werkzeuge (analog oder digital) operationalisiert.

Im Forschungsprojekt FLOW werden stets neue Techniken und Werkzeuge erarbeitet und evaluiert. Im Folgenden werden Techniken und zugehörige Werkzeuge dargestellt, die bisher existieren und sich zum Teil in der industriellen Praxis bewährt haben.

## Übersicht

| **Technik** | **Werkzeug** | **Aktivitäten je Phase** Erheben | Analysieren | Verbessern |
|---|---|---|---|---|
| Elicitation | Elicitation Package | Elicitation | | |
| (Kapitel 4.1, S. 25) | | Elicitation Packages in Interviews ausfüllen | - | - |
| Prozessmodell-ableitung | ADONIS-to-DocFLOW | Prozessmodellableitung | Prozessmodellableitung | |
| (Kapitel 4.2, S. 26) | | Transformation von ADONIS nach DocFLOW | (Semi-)automatische Identifikation von Problemen | - |
| Simulation | ProFLOW.swquant SWQuant-Spiel | Simulation | Simulation | |
| (Kapitel 4.3, S. 27) | | Modellierung mit SW-Quanten | Simulation der SW-Quanten | - |
| FLOW-Mapping | Web FLOW Map | FLOW-Mapping | FLOW-Mapping | FLOW-Mapping |
| (Kapitel 0, S. 29) | | Konnektoren für Kommunikations-kanäle einsetzen | Visualisierung von Kommunikations-events | Förderung gezielter Kommunikation |
| Schnittstellen-variationstechnik | Tailoring-Regeln | Schnittstellenvariationstechnik | Schnittstellenvariationstechnik | Schnittstellenvariationstechnik |
| (Kapitel 0, S. 35) | | Aktivität modellieren | FLOW-Interface analysieren | Interface variieren, Aktivität anpassen |
| SCRUM-Integration | V-Modell-to-FLOW, Integrationsleitfaden | SCRUM-Integration-in-V-Modell-XT | SCRUM-Integration-in-V-Modell-XT | SCRUM-Integration-in-V-Modell-XT |
| (Kapitel 0, S. 37) | | Informationsflüsse ableiten | Integrationspunkte identifizieren | SCRUM in V-Modell XT integrieren |
| FLOW-Muster | Musterkatalog | | FLOW-Muster | FLOW-Muster |
| (Kapitel 4.7, S. 39) | | | Mustersuche, Pattern-Matching | Muster gezielt einsetzen |
| Erfahrungs-verfestigung | LID-Template | | | Erfahrungs-verfestigung |
| (Kapitel 4.8, S. 40) | | - | - | LID-Sitzung durchführen |
| Anforderungs-verfestigung | FastFeedback | | | Anforderungs-verfestigung |
| (Kapitel 4.9, S. 41) | | - | - | Anforderungs-Interview |
| Prototyp-Demo-Verfestigung | FOCUS | | | Prototypdemo-verfestigung |
| (Kapitel 4.10, S. 42) | | - | - | Prototyp-Demonstration |
| Verfestigung als Nebenprodukt | Skype Solidifier | | | Meeting-verfestigung |
| (Kapitel 4.11, S. 43) | | - | - | verteilte Kommunik. |

Die folgenden Darstellungen der Techniken beginnen jeweils mit einer allgemeinen Beschreibung. Anschließend werden unterstützte Aktivitäten in den Phasen der FLOW-Methode erläutert oder es wird auf Literatur verwiesen, die eine detaillierte Beschreibung enthält. Falls vorhanden, werden Werkzeuge vorgestellt, die die Durchführung der Aktivitäten der FLOW-Techniken unterstützen. Eine Zusammenfassung beendet die Vorstellung jeder Technik. Dabei wird zu jeder FLOW-Technik eine kurze Schritt-für-Schritt-Anleitung des wesentlichen Vorgehens und eine Klassifikation entsprechend des Templates aus Tabelle 6 vorgestellt. Der Abgleich des Templates aus der Vorbereitungsphase mit den Einordnungs-Templates der Technikbeschreibungen erleichtert das Finden geeigneter Techniken für ein FLOW-Verbesserungsvorhaben (vgl. Kapitel 3.2).

## 4.1 Informationsflüsse Elicitieren

### Beschreibung

Bei der Elicitation geht es darum, durch aktives Befragen der am Informationsfluss beteiligten Personen ein Informationsflussmodell zu erheben. Dieses kann in späteren Phasen für Analyse und Verbesserung genutzt werden. Die Ergebnisse der Befragungen einzelner Personen werden in einem nachgelagerten Schritt zu einem Gesamtmodell zusammengefasst. Bei der Elicitation handelt es sich um eine Bottom-Up Strategie, da ausgehend von vielen unabhängigen Einzelbefragungen ein zusammenhängendes Gesamtmodell erstellt wird. Hauptverfahren der Elicitation ist das Interview. Die Strukturierung der Interviews wird durch Einsatz des Elicitation Packages erreicht. Das Elicitation Package besteht aus zwei Teilen: (1) einem Fragebogen und (2) einer FLOW-Aktivitätsschablone (siehe Anhang A). Die Fragen aus dem Fragebogen leiten das Interview an. Die Aktivitätsschablone erfasst die Informationsfluss-relevanten Eigenschaften der Hauptaufgaben der interviewten Personen. Für ein Interview werden also ein Fragebogen und mehrere Aktivitätsschablonen benötigt. Die Zusammenfassung der Aktivitäten ergibt dann das Gesamtmodell. Das Ergebnis ist ein Ist-Modell, da durch die Befragung nach tatsächlich ausgeführten Aufgaben auch die tatsächlichen Informationsflüsse erhoben werden.

### Zusammenfassung

Die 6 Schritte der Elicitations-Technik sind:

1. **Interviewees identifizieren:** Je nach Umfang des verfolgten Ziels (Aktivität, Projekt, Organisation) kommen mehr oder weniger Personen für ein Interview in Frage. Falls nicht genug Ressourcen für Interviews aller relevanten Personen zur Verfügung stehen, sollte man anhand von Vorwissen (z. B. Projektleiter befragen) problematische Stellen identifizieren und von dort gezielt zentrale Personen für die Interviews auswählen.

2. **Interviews planen und durchführen:** Wenn alle Interviewees identifiziert sind, werden Termine für die Interviews festgelegt. Im Interview werden mit Hilfe der anleitenden Fragen aus dem Elicitation-Fragebogen für jede wichtige Aktivität des Interviewees eine FLOW-Aktivitätsschablone ausgefüllt. Dabei ist es wichtig, dass alle für die Aktivität notwendigen Informationsspeicher und alle aus der Aktivität entstehenden Informationsspeicher korrekt aufgenommen werden, insbesondere auch die flüssigen.

3. **Aktivitätsschablonen in FLOW-Aktivitäten überführen:** Für jede in den Interviews ausgefüllte Aktivitätsschablone wird eine FLOW-Aktivität modelliert. Die Aktivität bekommt den Namen der Aufgabe. Eingehende, ausgehende, steuernde und kontrollierende Speicher werden im entsprechenden Aggregatzustand eingezeichnet.

4. **Anknüpfpunkte zwischen FLOW-Aktivitäten identifizieren:** Die schwierigste Aufgabe der Elicitationstechnik ist die Identifikation von Anknüpfpunkten zwischen den verschiedenen FLOW-Aktivitäten. Diese sind nötig um ein zusammenhängendes FLOW-Modell erstellen zu können (falls dieses existiert). Bei einer echten Zusammenarbeit sollten Verbindungen zwischen Aktivitäten in Form von Informationsspeichern mit dem gleichen bzw. ähnlichen Namen und dem gleichen Aggregatzustand existieren.

5. **Rückfragen an Interviewees bei unklaren Anknüpfpunkten:** Es kann vorkommen, dass bei der Suche nach Anknüpfpunkten weitere Fragen bezüglich der Namen oder des Aggregatzustandes der modellierten Informationsspeicher aufkommen. Diese sollten durch Rückfragen bei den beteiligten Personen geklärt werden, damit ein korrektes Modell entstehen kann.
6. **Gesamtmodell aus FLOW-Aktivitäten erstellen:** Abschließend sind alle FLOW-Aktivitäten in einem Gesamtmodell zusammenzufassen. Informationsspeicher, die als Anknüpfpunkte identifiziert wurden, kommen im Gesamtmodell nur einmal vor und stellen so die Verbindung zwischen den Aktivitäten her.

Das so erstellte Gesamtmodell steht nun den folgenden Phasen der FLOW-Methode zur Verfügung.

Die Technik **Informationsflüsse Elicitieren** verfolgt eine *Bottom-Up* Strategie, bei der *Ist*-Informationsflüsse mittels *Interviews* erhoben werden. Das zugehörige Elicitation Package kann in Einzelfällen auch durch direktes Ausfüllen der Entwickler (Fragebogen) oder als Anleitung beim Beobachten oder Selber-Machen durch einen FLOW-Experten genutzt werden.

**Tabelle 16. Einordnung der Technik "Informationsflüsse Elicitieren"**

**Informationsflüsse Elicitieren**

| | | | | | |
|---|---|---|---|---|---|
| **Phase** | | ☑ Erheben | ☐ Analysieren | ☐ Verbessern | |
| **Ziel** 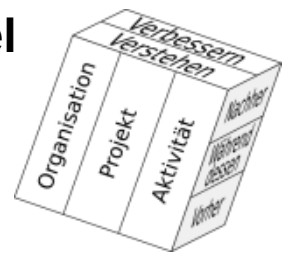 | **Absicht** | ☑ Verstehen | ☑ Verbessern | | |
| | **Zeit** | ☐ Vorher | ☑ Während dessen | ☑ Nachher | |
| | **Umfang** | ☑ Aktivität | ☑ Projekt | ☑ Organisation | |
| **Phasenspezifische Aspekte** | | | | | |
| **Erhebungsstrategie** | | ☑ Bottom-Up | ☐ Top-Down | | |
| **Erhebungsverfahren** | | ☐ Selber machen<br>☐ Modellableitung | ☐ Beobachten<br>☐ Kommunikationsereignisableitung | ☑ Interview | ☐ Fragebogen |

# 4.2 Informationsflüsse aus Prozessmodell ableiten

## Beschreibung

Bei der Technik „Informationsflüsse aus Prozessmodell ableiten" geht es darum, basierend auf vorhandenen Prozessmodellen ein Informationsflussmodell zu erstellen. Dies soll anschließend Analysen und Verbesserungen aus Informationsflusssicht ermöglichen, und somit die Prozesssicht erweitern. Bei der Prozessmodellableitung handelt es sich um eine Top-Down Erhebungsstrategie, da ausgehend von einem vorhandenen Gesamtmodell Informationsflüsse abgeleitet und ggf. Schritt für Schritt weiter verfeinert werden. In den meisten Prozessmodellen sind Dokumente die einzigen modellierten Informationsträger. Über die Zugehörigkeit der Dokumente zu Aktivitäten und den Kontrollflüssen zwischen den Aktivitäten kann in einem ersten Schritt ein Dokumentenflussmodell erstellt werden. Dies stellt eine Untermenge des Informationsflussmodells dar. Die flüssigen Informationsflüsse sind entweder schwer oder gar nicht aus dem Prozessmodell ableitbar. Ein Ansatzpunkt bieten die beteiligten und verantwortlichen Rollen der Aktivitäten und deren Beschreibungen. Um ein vollständigeres Informationsflussmodell zu bekommen wird man aber meist nicht um den zusätzlichen Einsatz einer anderen Erhebungstechnik herum kommen, wie zum Beispiel Interviews. Das Ergebnis der Ableitung aus einem Prozessmodell ist ein Soll-Modell, da modellierte Prozesse einen Soll-Zustand repräsentieren. Bei der späteren Analyse und Verbesserung ist zu beachten, dass dieser im Einzelfall sehr stark von den Ist-Zuständen abweichen kann.

Eine ausführliche Beschreibung zur Ableitung von Informationsflussmodellen aus Prozessmodellen befindet sich in der Masterarbeit von Kai Stapel:

> Kai Stapel, Informationsflussoptimierung eines Softwareentwicklungsprozesses aus der Bankenbranche (PDF − 1,1MB), Masterarbeit, Leibniz Universität Hannover, 2006

## Zusammenfassung

Zusammenfassend lässt sich das prinzipielle Vorgehen zur Ableitung von Informationsflüssen aus Prozessmodellen wie folgt beschreiben:

1. Für jedes Dokument über den Kontrollfluss einen Dokumentenfluss ableiten:
    a. Erste Referenz des Dokuments als Output einer Aktivität suchen. Diese Aktivität merken (Output-Aktivität).
    b. Dem Kontrollfluss folgen, bis eine Referenz als Input des Dokuments gefunden ist (Input-Aktivität).
    c. Es ergibt sich ein Dokumentenfluss zwischen den beiden Aktivitäten (Output-Aktivität → Input-Aktivität).
    d. Parallel zu b. und c. bei jeder untersuchten Aktivität den Output prüfen. Wird das Dokument dort gefunden, so wird diese Aktivität neue Output-Aktivität.
    e. Weiter bei b. bis das Ende des Prozessmodells erreicht ist.
2. Aus den einzelnen Modellen ein Gesamtdokumentenflussmodell erzeugen (über die Aktivitäten)
3. Ggf. über Rollenzuordnung flüssige Informationsflüsse ergänzen.
4. Ggf. durch Interviews vervollständigen (vgl. Kapitel 4.1).

Die Technik **Informationsflüsse aus Prozessmodell ableiten** verfolgt eine *Top-Down* Strategie, bei der *Soll*-Informationsflüsse aus bereits vorhandenen Prozessmodellen abgeleitet werden. Typischerweise geht es bei dieser Technik um organisationsweite Informationsflüsse. In Einzelfällen können damit auch projektspezifische Informationsflüsse erhoben werden.

**Tabelle 17. Einordnung der Technik "Informationsflüsse aus Prozessmodell ableiten"**

**Informationsflüsse aus Prozessmodell ableiten**

| | | | | | |
|---|---|---|---|---|---|
| **Phase** | | ☑ Erheben | ☑ Analysieren | ☐ Verbessern | |
| **Ziel** 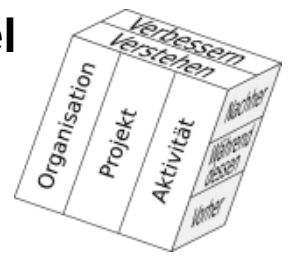 | **Absicht** | ☑ Verstehen | ☑ Verbessern | | |
| | **Zeit** | ☑ Vorher | ☐ Während dessen | ☐ Nachher | |
| | **Umfang** | ☐ Aktivität | ☑ Projekt | ☑ Organisation | |
| **Phasenspezifische Aspekte** | | | | | |
| **Erhebungsstrategie** | | ☐ Bottom-Up | ☑ Top-Down | | |
| **Erhebungsverfahren** | | ☐ Selber machen<br>☑ Modellableitung | ☐ Beobachten<br>☐ Kommunikationsereignisableitung | ☐ Interview | ☐ Fragebogen |

# 4.3 Simulation von Informationsflüssen

## Beschreibung

Bei der Simulation geht es darum, quantitatives Verhalten von Informationsflüssen in Softwareprojekten zu analysieren. Dazu bedarf es einer Informationseinheit, die Informationsinhalte auf ein simulierbares Maß abstrahiert. Eine in FLOW erprobte Einheit sind die Software-Quanten. Sie basieren auf der Software-Quanten-Metapher.

## Software-Quanten-Metapher

Die Software-Quanten-Metapher ist ein deskriptives Modell der Softwareentwicklung. Sie dient dazu Ist-Informationsflüsse quantitativ darzustellen, sodass man sowohl die Stärken als auch die Schwächen eines realen Prozesses sehen kann. Softwarequanten ermöglichen neben den quantitativen auch qualitative Aussagen. Das Modell wurde für folgende Modellierungszwecke gebildet:

- **Hauptziel:** Veranschaulichung von qualitativ-quantitativen Aspekten von Informationsflüssen in der Softwareentwicklung, um Fehler erkennen und zukünftig vermeiden zu können.
- Die dargestellten Aspekte sollen einfach verständlich und einprägsam sein.

- Neue Situationen sollen durch einfache Variationen des Modells ausprobiert und veranschaulicht werden können.
- Das Modell soll für animierte Simulationen geeignet sein (Simulation und Visualisierung).

Softwarequanten sind atomare Projektinformationseinheiten. Sie haben ungefähr ein Zehntel der Größe eines Function Points [11]:

Kurt Schneider: *A Descriptive Model of Software Development to Guide Process Improvement*. Conquest 2004

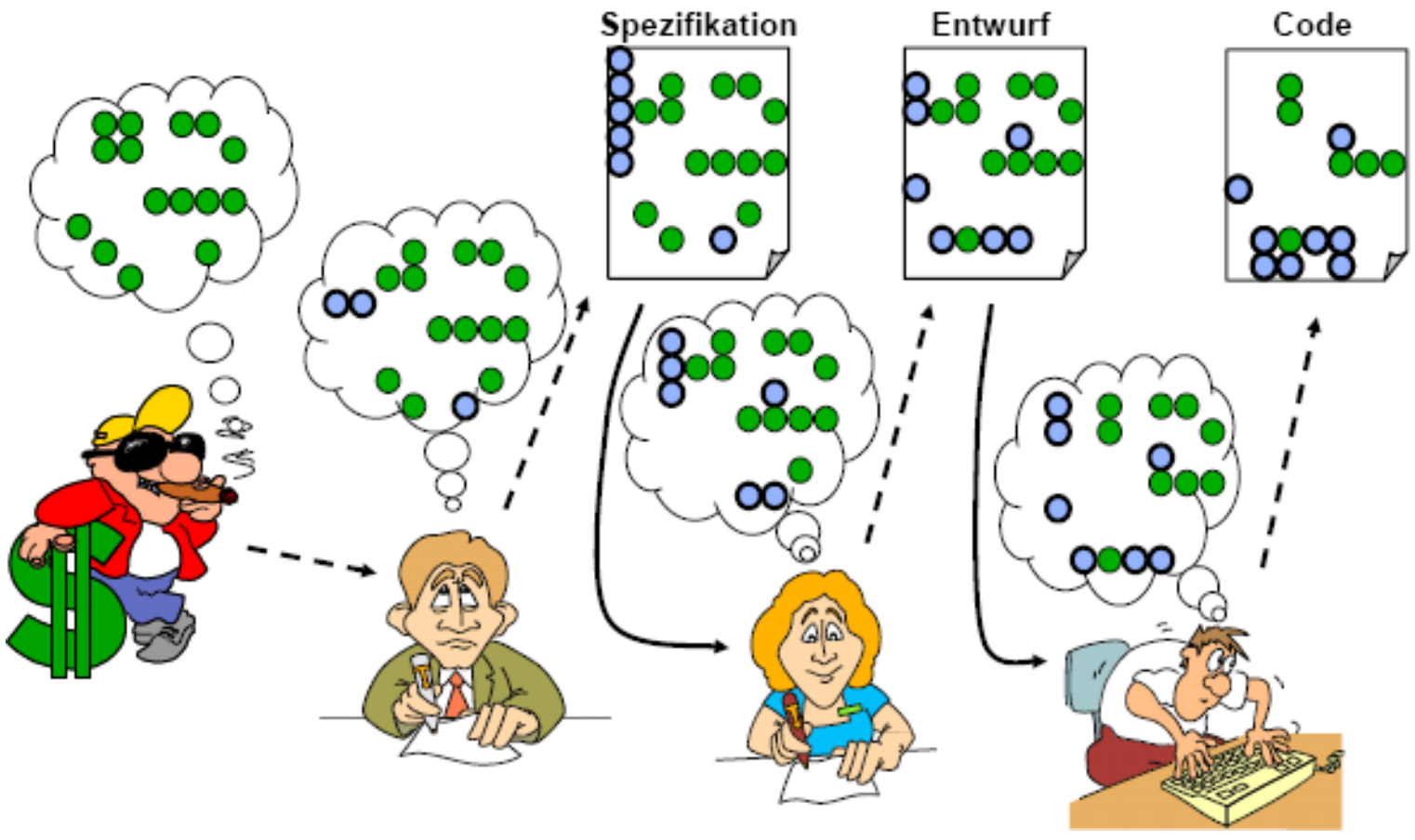

**Abbildung 8. Beispiel-Informationsfluss dargestellt mit der Software-Quanten-Metapher.**

Die Metapher bildet alle Anforderungen zu Beginn eines Projekts auf eine Menge von Kugeln ab. Die Kugeln und deren Anordnung stellen die Visualisierung der Softwarequanten dar. Die Quanten im Kopf des Kunden sind alle in der gleichen Farbe dargestellt (grüne Kugeln), da sie echte legitime Anforderungen sind. Die Anforderungen des Kunden bilden die Referenz. Wenn ein Analyst den Kunden interviewt, greift er sich eine (Unter-)Menge von Softwarequanten aus der Menge der Quanten des Kunden und fügt sie seiner eigenen Menge hinzu. Dort vermischen sie sich mit schon vorhandenen Softwarequanten, die eventuell aus vorherigen Projekten im gleichen Kontext stammen. Durch Missverständnisse können Quanten weggelassen oder hinzugefügt werden. Je länger ein Anforderungsinterview dauert, desto mehr Quanten können transferiert werden. Dabei ist es sehr wahrscheinlich, dass Anforderungen wiederholt übertragen werden. Die Anzahl der Quanten des Analysten erhöht sich dann nicht weiter. Richtig übertragene Quanten werden in derselben Farbe, wie die des Kunden, dargestellt (grün), falsche Quanten in einer anderen Farbe (blau). Dabei bestimmt sich die Korrektheit eines Quants immer in Bezug auf die Quanten des Kunden. Daher kann es zum Beispiel passieren, dass der Programmierer den Entwurf falsch versteht und etwas anderes als beschrieben programmiert, damit aber eine tatsächliche Anforderung des Kunden erfüllt, die auf dem Weg bis zum Entwurf schon vergessen war.

Mathematisch ist die Übertragung von Softwarequanten eine zufällige Auswahl aus einer Menge von unterscheidbaren Objekten ohne Beachtung der Reihenfolge mit Zurücklegen (Repetition, der Kunde vergisst seine Anforderungen für gewöhnlich nicht). Quanten sind individuell. In der Visualisierung werden gleiche Quanten durch die gleiche Position dargestellt.

Die Simulation von Informationsflüssen mit Hilfe der Software-Quanten-Metapher ist eine FLOW-Praktik, die schon zweimal angewandt wurde. Karsten Möckel hat in seiner Bachelorarbeit eine Simulationskomponente für das ProFLOW-Framework entwickelt:

Karsten Möckel, Simulationskomponente für Informationsflüsse in einem Prozessmodellierungs-Framework (PDF – 1,1MB), Bachelorarbeit, Leibniz Universität Hannover, 2007

Im Softwareprojekt 2007/2008 haben Studenten ein Software-Quanten-Spiel implementiert, das die Wichtigkeit von Requirements Engineering in der Softwareentwicklung spielerisch veranschaulicht:

Eric Knauss, Kurt Schneider, Kai Stapel: *A Game for Taking Requirements Engineering More Seriously*. Third International Workshop on Multimedia and Enjoyable Requirements Engineering (MERE 08), Barcelona, Spain, 2008

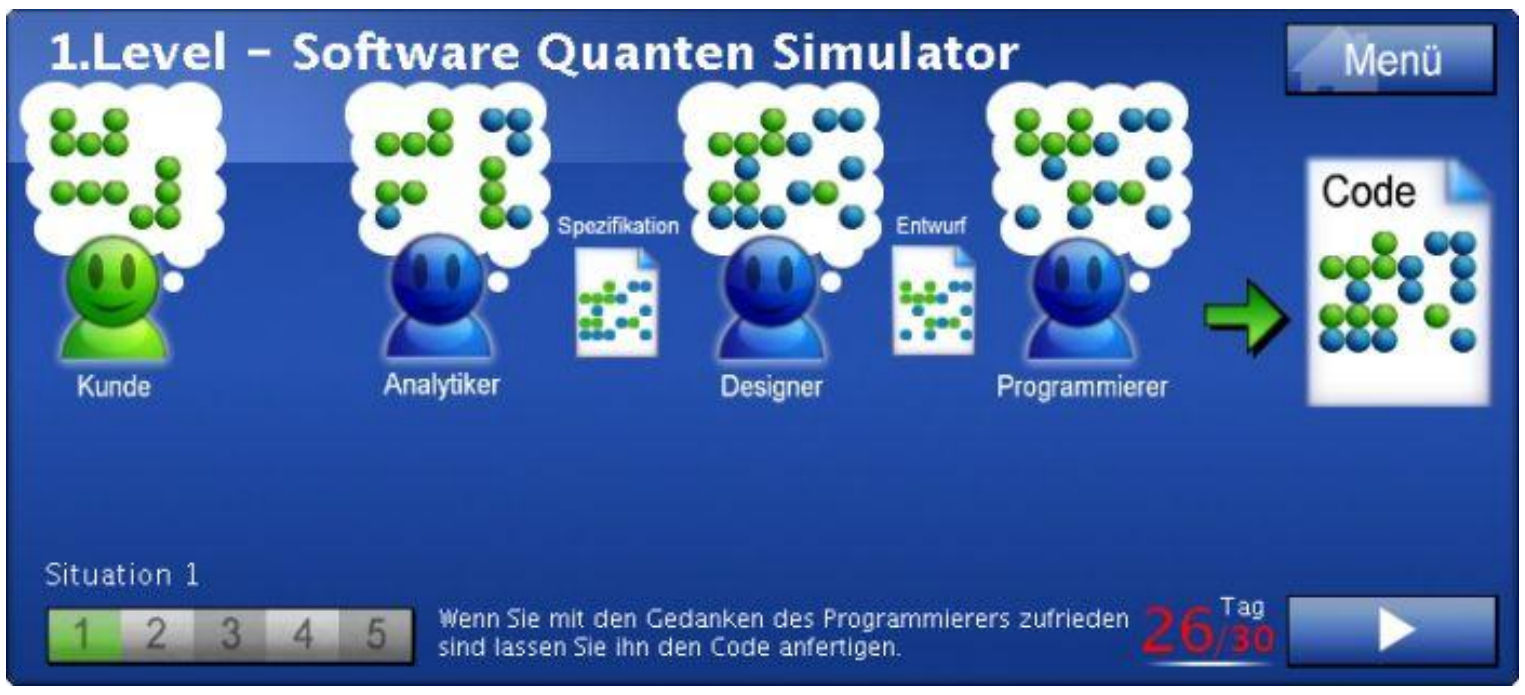

**Abbildung 9. Screenshot des Software-Quanten-Spiels**

## Zusammenfassung

Zusammengefasst gibt es folgendes Vorgehen bei der Simulation von Informationsflüssen:

1. Zu simulierende Informationsflusseigenschaft auswählen (hier Softwarequanten)
2. Simulationsmodell erstellen (Bsp. bei Kommunikation und Dokumentation können Informationen, d.h. Softwarequanten, verfälscht werden)
3. Simulator implementieren
4. Simulation justieren, d.h. Simulationsparameter einstellen (z.B. die Verfälschungsrate)
5. Verschiedene Entwicklungssituation mit Hilfe der Simulation analysieren

Die Technik der **Simulation** von Informationsflüssen ist ein Simulationsverfahren, das auch durch Visualisierung der Simulationsergebnisse oder Pattern-Matching-Verfahren erweitert werden kann. Die Simulation dient vornehmlich dem Verständnis und bietet daher nur Aktivitäten für die Phasen *Erheben* und *Analysieren*.

**Tabelle 18. Einordnung der Technik "Simulation"**

| | | Simulation | | |
|---|---|---|---|---|
| **Phase** | | ☑ Erheben | ☑ Analysieren | ☐ Verbessern |
| **Ziel** 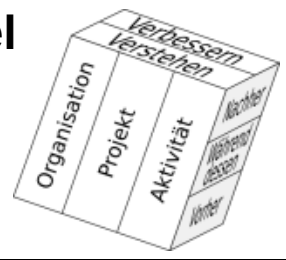 | **Absicht** | ☑ Verstehen | ☐ Verbessern | |
| | **Zeit** | ☐ Vorher | ☑ Während dessen | ☐ Nachher |
| | **Umfang** | ☐ Aktivität | ☑ Projekt | ☐ Organisation |
| **Phasenspezifische Aspekte** | | | | |
| **Verfahren** | | ☐ Visualisierung | ☑ Simulation | ☐ Pattern Matching |

# 4.4 FLOW-Mapping

## Beschreibung

Mit der FLOW-Mapping-Technik werden Soll- und Ist-Informationsflüsse visualisiert, mit einem Fokus auf flüssige Flüsse. Die Visualisierung soll die Awareness im Entwicklungsteam verbessern. Dies ist insbesondere in verteilten Teams wichtig, wo sich nicht alle Projektbeteiligten kennen, nicht wissen wer was weiß oder wer woran arbeitet. Erhöhte Awareness im Team und die zusätzliche Darstellung von Kontaktinformationen soll schließlich gezielte Kommunikation fördern und somit die Kommunikation im Projekt verbessern.

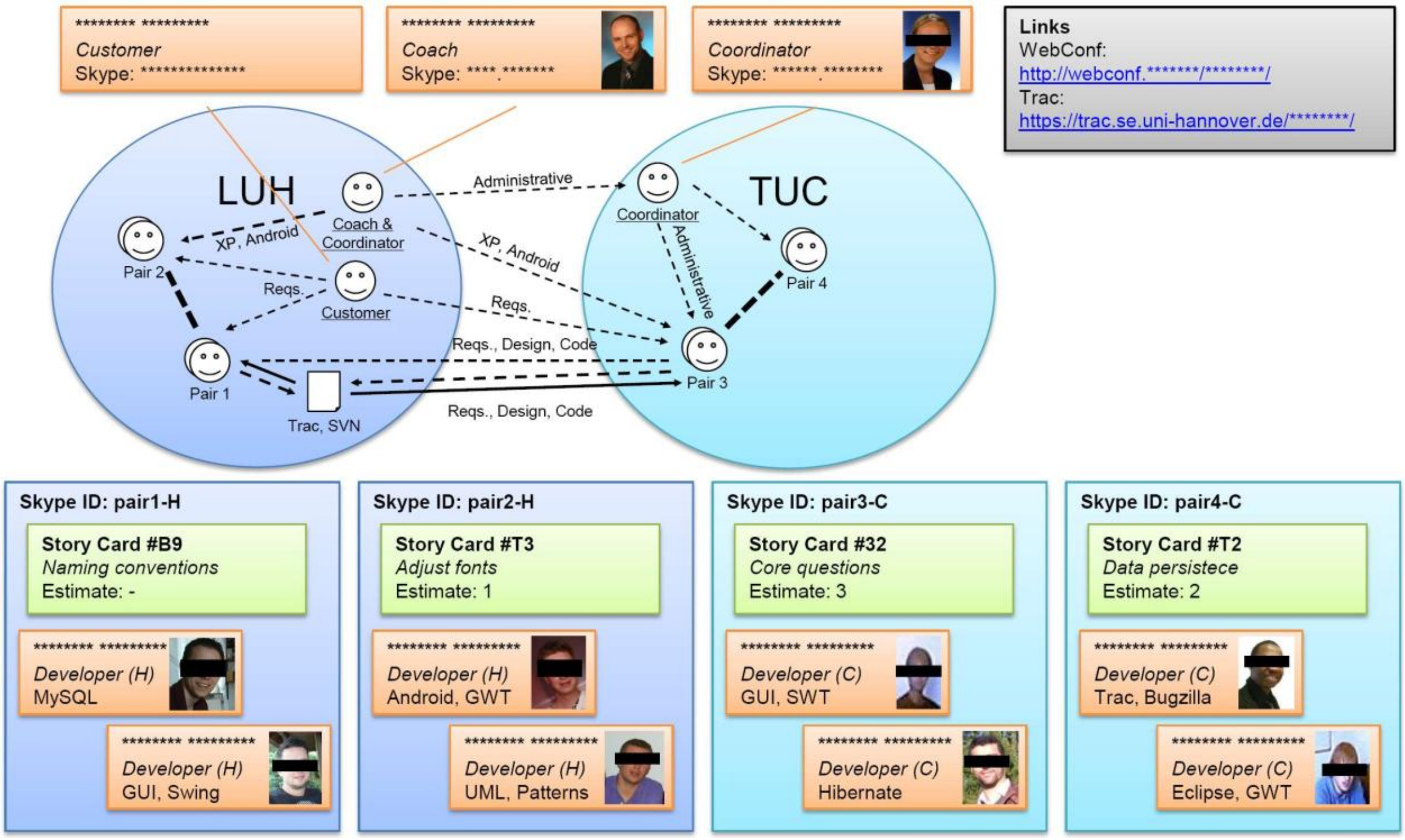

**Abbildung 10. Beispiel FLOW-Map eines verteilten Projekts.**

## FLOW-Map

Kern und namensgebender Bestandteil der FLOW-Mapping-Technik ist die so genannte FLOW-Map. Eine FLOW-Map ist ein FLOW-Modell, welches eine erweiterte FLOW-Notation nutzt, um Spezifika verteilter Softwareprojekte darstellen zu können. Eine FLOW-Map hat folgende zusätzliche Bestandteile:

- Zuordnung von Informationsspeichern zu Entwicklungsstandorten. Der Ort eines flüssigen Speichers repräsentiert den physischen Standort der Person, an die die flüssige Information gebunden ist. Der Ort eines festen Speichers repräsentiert den Standort welcher für den Inhalt des Speichers verantwortlich ist, d.h. insbesondere nicht seinen physischen Standort.
- Unterschiedliche Liniendicken zur Unterscheidung unterschiedlich intensiver Informationsflüsse.
- Ungerichtete Flüsse zur Darstellung von Informationsflüssen in beide Richtungen.
- Optional: Piktogramme an standortübergreifenden Informationsflüssen zur Darstellung der genutzten Kommunikationsmedien.
- Zusätzliche Meta-Daten für jeden Speicher, die so genannten "Gelben Seiten": Kontaktinformationen (Name, E-Mail-Adresse, etc.), Bilder der Entwickler, lokale Zeit, Statusinformationen (beschäftigt, verfügbar), Rolle im Projekt, projektrelevante Fähigkeiten, aktuelle Aufgabe (z.B. bestimmtes Bug-Ticket oder User Story) und aktueller Arbeitsgegenstand (z.B. bestimmte Teile des Quellcodes), usw.
- Erweiterung für agile Projekte: Ein doppeltes Smiley-Symbol zur Darstellung von Pair-Programmierern
- Erweiterung für agile Projekte: Erweiterte Gelbe-Seiten-Informationen zur Darstellung der Zusammensetzung von Programmierpaaren.

Eine FLOW-Map visualisiert Projektbeteiligte als flüssige Speicher, ihren Entwicklungsstandort, zentrale standortübergreifende Dokumente als feste Speicher, Gelbe-Seiten-Informationen und geplante bzw. tatsächliche Informationsflüsse zwischen allen Speichern. Ein Beispiel einer FLOW-Map mit geplanten Informationsflüssen und Gelbe-Seiten-Informationen ist in Abbildung 10 dargestellt.

Eine FLOW-Map kann neben der Awarenesssteigerung auch als kognitive Hilfe sowohl bei der Planung als auch bei der Durchführung der Kommunikation in verteilten Projekten dienen. Im Folgenden wird anhand der drei Phasen des FLOW-Verbesserungsprozesses beschrieben, wie mit Hilfe der FLOW-Map die Kommunikation in verteilten Projekten geplant und durchgeführt werden kann.

## Phase 1: Erheben

FLOW-Mapping bietet in der ersten Phase des FLOW-Verbesserungsprozesses Unterstützung für die Planung der Kommunikation in einem verteilten Projekt und Unterstützung zur Ableitung der Informationsflüsse mit Hilfe von Kommunikationsereignissen. Abbildung 11 zeigt einen Überblick über die Aktivitäten der ersten Phase in FLOW-Notation. Zunächst werden die zentralen Rollen in FLOW-Mapping beschrieben.

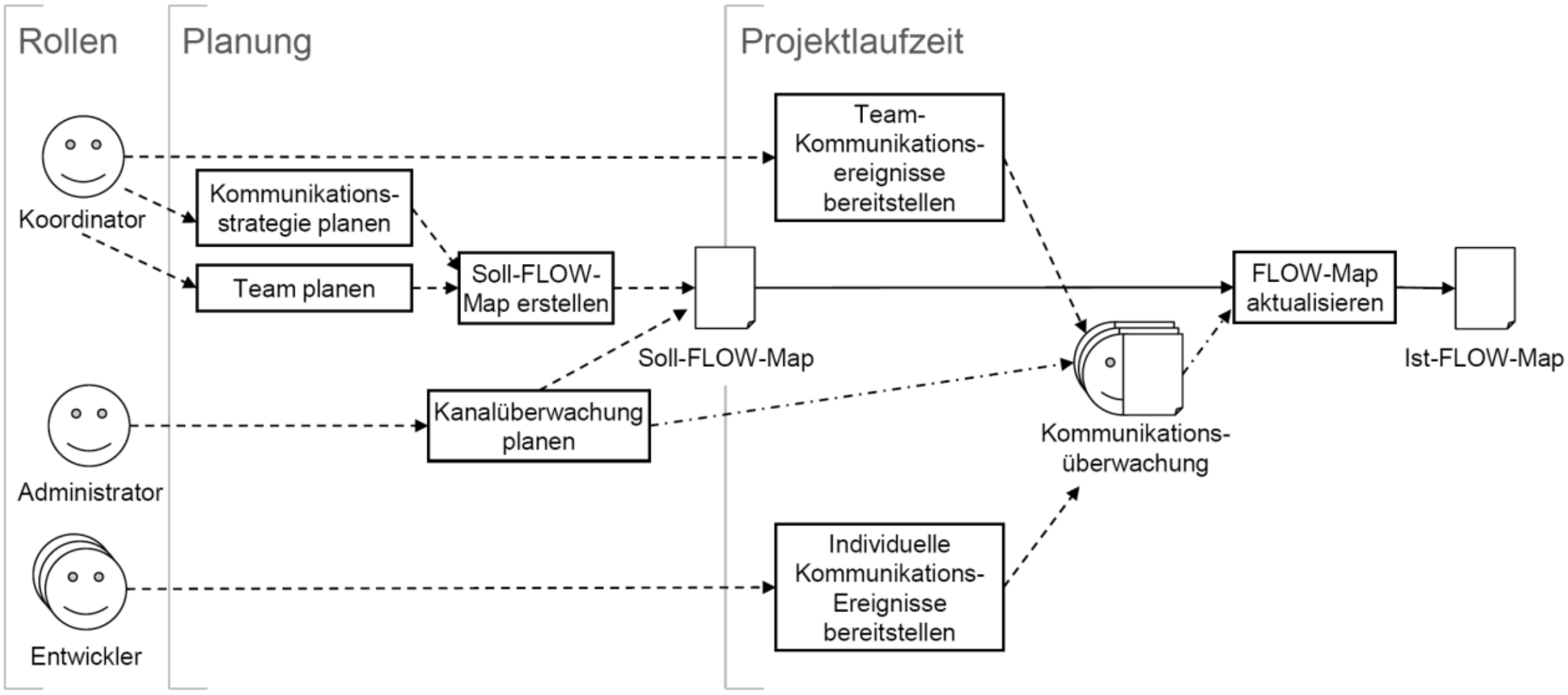

**Abbildung 11. FLOW-Mapping Aktivität der Phase 1: Erheben**

### *Rollen in FLOW-Mapping*

Bei der Durchführung der FLOW-Mapping-Technik werden drei Rollen unterschieden:

1. **Koordinator:** Die koordinierende Rolle, z.B. ein Projektleiter oder Kommunikations-Coach, übernimmt planende, überwachende und steuernde Aufgaben bei der Durchführung von FLOW-Mapping. Der Koordinator ist für die Informationsflüsse des gesamten Projekts verantwortlich.
2. **Entwickler:** Entwickler nutzen die FLOW-Map zur Steigerung der Team-Awareness und zur Initiierung gezielter Kommunikation. Sie liefern individuelle Kommunikationsereignisse zur Steigerung der Awareness ihrer Kollegen und damit der Koordinator einen Überblick gewinnen kann.
3. **Administrator:** Die administrative Rolle übernimmt Aufgaben zur Bereitstellung und Aktualisierung der FLOW-Map.

Eine Person kann durchaus auch mehrere Rollen übernehmen, z.B. die Koordination und die Administration.

### *Planung der Kommunikation*

Wie in Abbildung 11 gezeigt, besteht die Planung der Kommunikation aus vier Aktivitäten. Der Koordinator plant zunächst das Team und die Kommunikationsstrategie (vgl. auch [20]). Beides ist Voraussetzung für die Erstellung der Soll-FLOW-Map für das Projekt. Für die in der Kommunikationsstrategie geplanten Medien bereitet der Administrator die Kanalüberwachung vor. Nach der Planung werden die Soll-FLOW-Map und Informationen der Vorbereitung der Kanalüberwachung im laufenden Projekt benutzt (siehe unten).

Bei der Planung des Teams werden die beteiligten Standorte, Entwickler im verteilten Team und andere für die Kommunikation wichtige Dinge, wie die für die standortübergreifende Kommunikation zu nutzende Sprache und Mindestanforderungen an Programmierfähigkeiten, festgelegt. Schließlich müssen Kontaktdaten aller Projektbeteiligten gesammelt werden.

Eine Kommunikationsstrategie ist eine Menge von geplanten oder Ereignis-basierten Kommunikationsaktivitäten mit zugehörigen zu nutzenden Kommunikationsmedien. Typische Kommunikationsaktivitäten in verteilten Projekten sind zum Beispiel tägliche verteilte Meetings zur Abstimmung oder ad-hoc Abstimmung zwischen zwei Entwicklern, die gemeinsam an einem Problem arbeiten. Die Kommunikationsstrategie wird in drei Schritten erstellt:

1. **Kommunikationsaktivitäten planen:** In diesem Schritt werden die Kommunikationsaktivitäten festgelegt, die für die weitere Betrachtung (Medienplanung und Überwachung) wichtig sind. Dabei werden zwei Arten von Aktivitäten unterschieden: (1) Aktivitäten, die regelmäßig, z.B. jeden Morgen oder zum Ende einer Woche, durchgeführt werden sollen, wie tägliche Stand-Ups, und (2) Aktivitäten, die zu einem bestimmten Ereignis ausgeführt werden sollen, wie Ad-hoc-Kommunikation, wenn eine Frage aufkommt, oder das Weiterleiten von Änderungswünschen des Kunden. Kommunikationsaktivitäten können aus dem verwendeten Softwareentwicklungsprozess oder aus Erfahrung aus ähnlichen Projekten abgeleitet werden.
2. **Mediennutzung planen:** Für jede Kommunikationsaktivität werden passende Kommunikationsmedien festgelegt, damit die Kanalüberwachung vorbereitet werden kann und während des Projekts der Aufgabe angemessene Medien genutzt werden. Für die Wahl geeigneter Medien können Theorien zur Medienwahl wie die Media Synchronicity Theory [2] genutzt werden. Bei verteilter Entwicklung ist es besonders wichtig, dass gezielt ein Kommunikationskanal für die Inhaltsübermittlung und ein Kanal für die Steuerung gewählt werden, statt nur einen Kanal für beides einzuplanen.
3. **Kommunikationsaktivitäten-spezifische FLOW-Maps erstellen**: Der letzte Schritt bei der Planung der Kommunikationsstrategie ist die Erstellung spezifischer FLOW-Maps für jede Kommunikationsaktivität. Diese spezifischen FLOW-Maps helfen Teilnehmer (Personen und Standorte), Informationsflussbesonderheiten der jeweiligen Aktivität und zwischen den Standorten zu nutzende Medien darzustellen. Wie eine FLOW-Map erstellt werden kann wird im folgenden Abschnitt beschrieben.

Mit Hilfe der Ergebnisse der vorangegangenen Aktivitäten, d.h. mit den Informationen über das Team und der Kommunikationsstrategie, kann die Soll-FLOW-Map erstellt werden. Die Soll-FLOW-Map gibt die geplanten Informationsflüsse für ein Projekt wieder. Es wird festgelegt welche Informationsspeicher (Personen und Dokumente), über welche Medien, wie intensiv kommunizieren sollen. Die FLOW-Map kann mit den folgenden sechs Schritten erstellt werden:

1. Erstelle eigene Bereiche für jeden beteiligten Standort.
2. Erstelle einen flüssigen Speicher für jeden Projektbeteiligten. Ordne die Speicher ihren Standorten zu.
3. Erstelle einen festen Speicher für jedes Dokument bzw. jeden Datenspeicher, der standortübergreifend-relevante Informationen enthält.
4. Erstelle flüssige Informationsflüsse zwischen flüssigen Speichern, die Informationen regelmäßig austauschen sollen. Erstelle flüssige Informationsflüsse zwischen flüssigen Speichern und festen Speichern, wenn Informationen in Dokumenten verfestigt werden sollen. Wenn Informationen vorwiegend in eine Richtung fließen sollen, dann kann man das mit einer Pfeilspitze in die entsprechende Richtung markieren. Ansonsten werden ungerichtete Informationsflüsse notiert. Die Intensität des gewünschten Flusses kann durch die Liniendicke verdeutlicht werden. Diese ist auch abhängig vom verwendeten Kommunikationskanal. Zum Beispiel kann in einem lokalen Gespräch von Angesicht zu Angesicht viel mehr Information ausgetauscht werden, als in der gleichen Zeit über einen Sofortnachrichtendienst (vgl. z.B. gesprochen vs. geschrieben).
5. Erstelle feste Informationsflüsse, wenn Dokumente regelmäßig gelesen werden sollen.
6. Abschließend wird die FLOW-Map um folgende Zusatzinformationen zu den Projektbeteiligten ergänzt:
    - Porträt-Bilder
    - Kontaktinformationen, wie E-Mail-Adresse, Telefonnummer oder Skype-ID entsprechend den geplanten Kommunikationsmedien
    - Geplante Rolle im Projekt
    - Projektrelevante Fähigkeiten
    - Zu bearbeitende Aufgabe(n)

Die Soll-FLOW-Map kann während der Planung genutzt werden, um den aktuellen Stand der Planung zwischen Koordinatoren und Entwicklern zu kommunizieren und während des Projekts, um den Entwicklern eines Standorts zu zeigen, wer noch am Projekt beteiligt ist und wie diese erreicht werden können. Zudem geben die Liniendicken der Informationsflüsse die Intensität der geplanten Kommunikation an, sodass die Entwickler immer sehen können, mit wem sie regelmäßig in Kontakt sein sollten.

Die letzte Aktivität der Planung ist die Vorbereitung der Kanalüberwachung. Dies ist notwendig, damit später in der Ist-FLOW-Map ohne großen Aufwand, d.h. möglichst automatisch, aktuelle Kommunikationsereignisse visualisiert und Abweichungen vom Plan festgestellt werden können. Ein Vorteil verteilter Entwicklung ist, dass meist elektronische Medien zur standortübergreifenden Kommunikation genutzt werden. Diese können leichter automatisch überwacht werden als analoge Medien wie Briefpost oder lokale Meetings von Angesicht zu Angesicht. Für die Kanalüberwachung muss je Kanal ein Mechanismus geschaffen werden, der Kommunikationsereignisse erfassen und für die Visualisierung in der Ist-FLOW-Map verfügbar machen kann. Das kann z.B. eine Software sein, die Start, Ende und Teilnehmer einer Skype-Telefonkonferenz aufzeichnet und weiterreicht. Für Kanäle, die gar nicht oder nur mit sehr hohem Aufwand automatisch überwacht werden könnten, können auch Personen die Aufgabe der Erfassung und Weiterleitung von Kommunikationsereignissen übernehmen. Einige Beispiele für Werkzeuge zur Kanalüberwachung sind in [20] beschrieben. Als Hilfsmittel zur Vorbereitung für den späteren Soll-Ist-Vergleich sei auf die für Kommunikationsaktivitäten angepassten Conformance-Templates nach Zazworka verwiesen [22].

> Nico Zazworka, Kai Stapel, Eric Knauss, Forrest Shull, Victor R. Basili, Kurt Schneider: Are Developers Complying with the Process: An XP Study, Proceedings of the 4th International Symposium on Empirical Software Engineering and Measurement (ESEM '10), 2010.
>
> Kai Stapel, Eric Knauss, Kurt Schneider, Nico Zazworka: FLOW Mapping: Planning and Managing Communication in Distributed Teams, Proceedings of 6th IEEE International Conference on Global Software Engineering (ICGSE '11), 2011.

### *Projektdurchführung*

Nachdem die Kommunikation mit Hilfe von FLOW-Mapping geplant ist und die Soll-FLOW-Map erstellt wurde, kann das Projekt starten. Für eine effektive Awarenesssteigerung sollten die Projektbeteiligten nicht nur die geplanten, sondern auch die tatsächlichen Informationsflüsse sehen können. Daher werden während des Projekts mit Hilfe der Mechanismen der Kommunikationsüberwachung aktuelle Kommunikationsereignisse erfasst und in einer Ist-FLOW-Map dargestellt (vgl. Abbildung 11, rechts).

Die Kommunikationsüberwachung sammelt Kommunikationsereignisse automatisch oder manuell, bei denen das gesamte Team oder nur ein Teil des Teams beteiligt sind, und stellt sie für die Aktualisierung der FLOW-Map zur Verfügung. Bei der Aktualisierung der Ist-FLOW-Map sollten folgende Änderungen berücksichtigt werden:

- Aktuelle Kommunikationsereignisse, wie Team-Meetings, Telefonkonferenzen, Dokumentenänderungen (z.B. Wiki-Änderungen) oder Quellcodeänderungen (z.B. via Versionsverwaltungssystem).
- Änderungen der Gelbe-Seiten-Informationen, z.B. wenn sich Rollen, Aufgaben oder Teamzusammensetzungen ändern.

Insgesamt wird für die Erhebung der Informationsflüsse beim FLOW-Mapping eine Bottom-Up-Strategie und das Verfahren der Kommunikationsereignisableitung genutzt (vgl. Kapitel 3.3). Weitere Ausführungen, wie FLOW-Mapping zur Planung der Kommunikation in verteilten Projekten genutzt werden kann, finden sich in:

> Kai Stapel, Eric Knauss, Kurt Schneider, Nico Zazworka: FLOW Mapping: Planning and Managing Communication in Distributed Teams, 6th IEEE International Conference on Global Software Engineering (ICGSE '11), 2011.

## Phase 2 & 3: Analysieren & Verbessern

Zur Analyse der Kommunikation nutzen die Entwickler und der Koordinator die Ist-FLOW-Map (vgl. Abbildung 12). Wenn die Ist-FLOW-Map die aktuellen Informationsflüsse enthält, können die Entwickler sehen, wer mit wem kommuniziert, wer im Moment für Kommunikation verfügbar ist und wer gerade an welcher Aufgabe arbeitet. Mit einem Vergleich von Soll- und Ist-FLOW-Map können Un-terschiede zur geplanten Kommunikation festgestellt werden. Auf Basis dieser Informationen können die Entwickler gezielter Kommunikation initiieren und der Koordinator die Kommunikation des gesamten Teams besser steuern. Beides führt zu einem verbesserten Informationsfluss im Projekt.

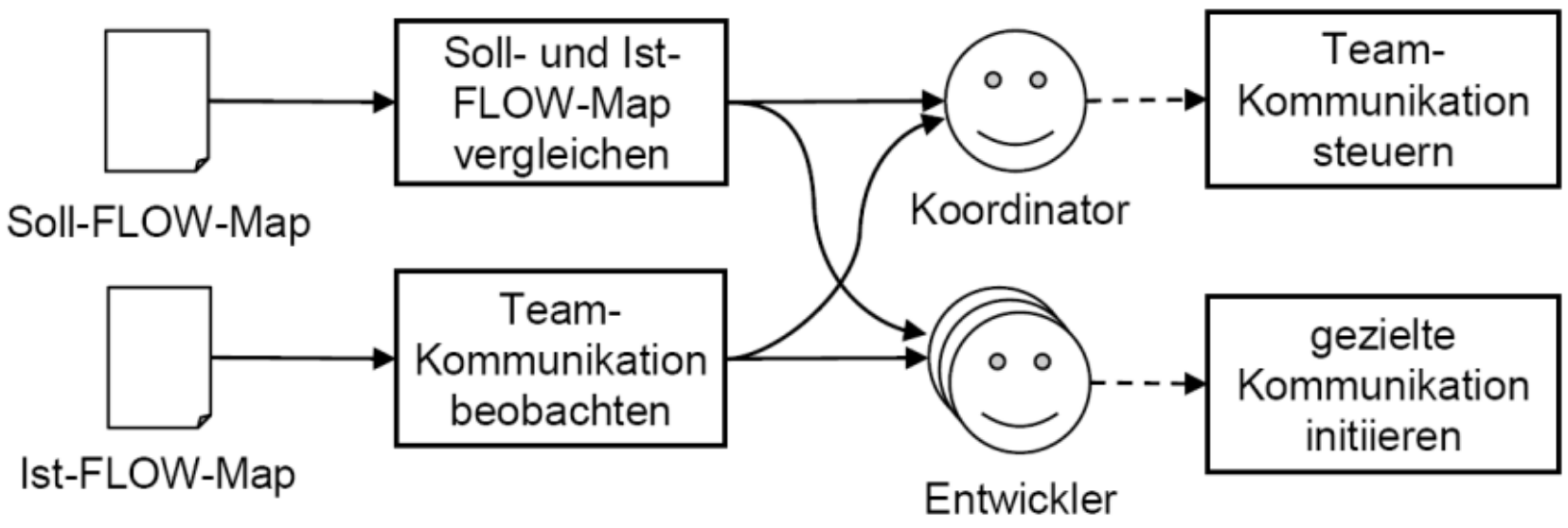

**Abbildung 12. FLOW-Mapping Aktivitäten der Phasen 2 & 3: Analysieren und Verbessern**

## Web FLOW Map

Die Web-FLOW-Map ist ein webbasiertes Werkzeug, das Unterstützung für folgende Aktivitäten der FLOW-Mapping-Technik bietet:

- Ist-FLOW-Map als Webseite darstellen
- Team-Kommunikationsereignisse entgegennehmen
- Individuelle Kommunikationsereignisse entgegennehmen
- Automatische Kanalüberwachung für Skype (Text, Telefon und Video), Subversion und Trac
- Aktuelle Ereignisse auf FLOW-Map visualisieren (Live-View)
- Unterstützung zur Initiierung gezielter Kommunikation
- Vergangene Ereignisse auf FLOW-Map visualisieren (History-View)
- Unterstützung bei Analyse des Kommunikationsverhaltens

Abbildung 13 zeigt einen Screenshot der Weboberfläche des Werkzeugs. Zentraler Bestandteil ist die Darstellung der FLOW-Map. Neben Standorten und wichtigen Informationsspeichern zeigt sie abhängig vom gewählten Modus entweder aktuelle Kommunikationsereignisse (Live-View) oder eine Zusammenfassung vergangener Kommunikationsereignisse (History-View). In der Live-View werden aktive Kommunikationsaktivitäten, die mehr als zwei Personen involvieren und aktive Gruppenchats unterhalb der FLOW-Map dargestellt. Dies geschieht außerhalb der FLOW-Map, um diese nicht visuell zu überladen. Auf der linken Seite gibt es Gelbe-Seiten-Informationen über den aktuell ausgewählten Entwickler und die Möglichkeit direkte Kommunikation mit diesem zu initiieren. Die Web-FLOW-Map wurde im Rahmen der Bachelorarbeit von Cieśnik entwickelt. Dort finden sich auch technische Details zur Implementierung.

> Robert Cieśnik, *Dashboard zur Verbesserung der Kommunikation bei verteilter Softwareentwicklung*, Bachelorarbeit, Leibniz Universität Hannover, 2010

## Zusammenfassung

Die folgenden Schritte fassen das Vorgehen von FLOW-Mapping grob zusammen:

1. Kommunikation für verteiltes Projekt planen
    a. Team zusammenstellen
    b. Kommunikationsstrategie erstellen
        i. Kommunikationsaktivitäten planen
        ii. Mediennutzung planen
        iii. Aktivitäts-spezifische FLOW-Maps erstellen (vgl [20])
    c. Soll-FLOW-Map erstellen
    d. Kanalüberwachung vorbereiten
2. Kommunikation während des Projekts steuern
    a. Kommunikation überwachen
    b. Ist-FLOW-Map aktualisieren
    c. Soll-Ist-Vergleich durchführen
    d. Kommunikation initiieren

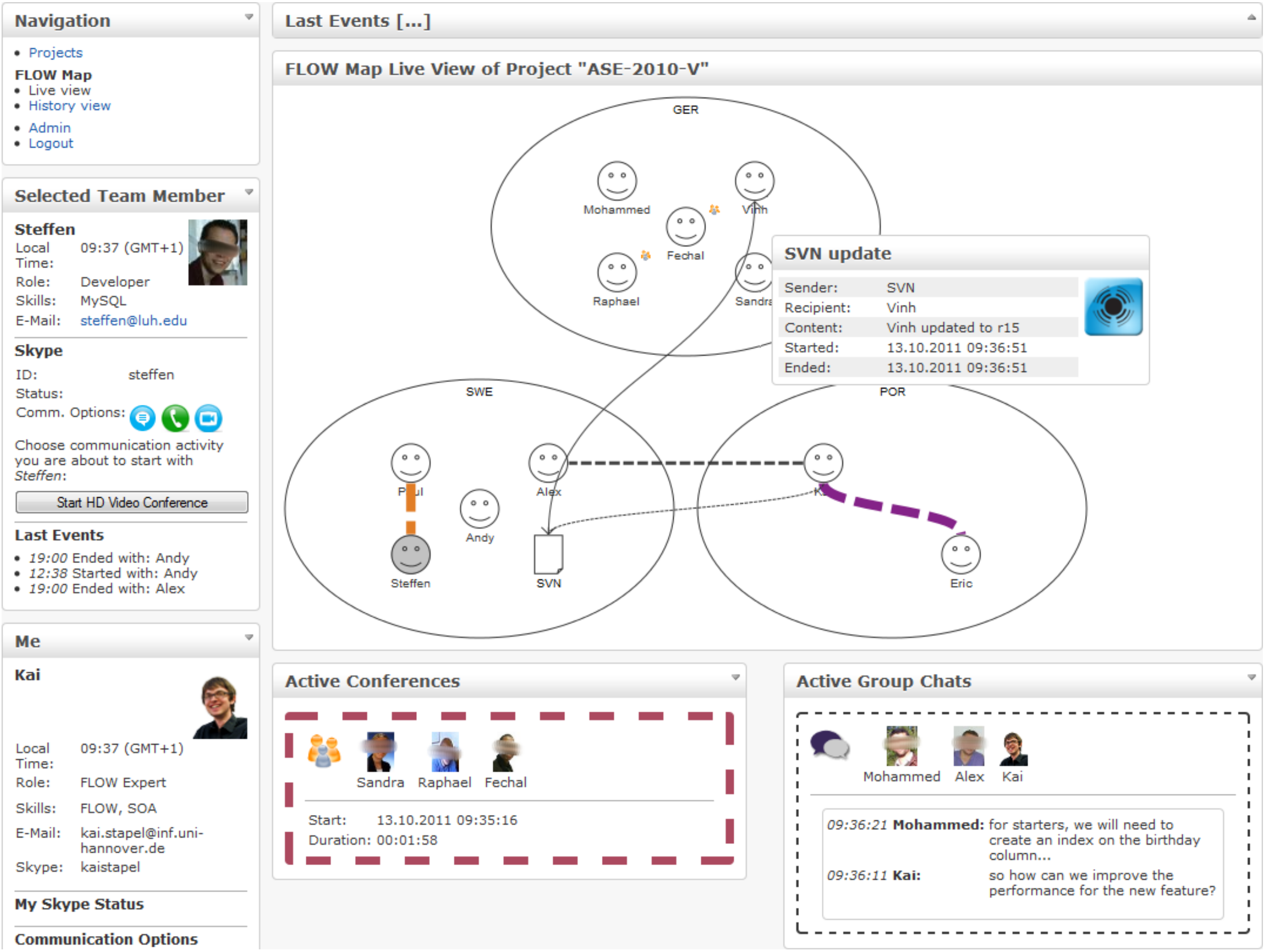

**Abbildung 13. Screenshot der Web-FLOW-Map**

**FLOW-Mapping** ist eine Technik zur *Analyse* und *Verbesserung* von Informationsflüssen *während* eines laufenden *Projekts*. Sie hat das Ziel insbesondere die flüssigen Informationsflüsse und -speicher eines Projekts *zu verstehen*, indem die wichtigsten flüssigen Informationsflüsse und -speicher durch *Visualisierung* allen Projektbeteiligten bewusst gemacht werden (Awareness-Steigerung). Dieses Bewusstsein soll im Team dafür sorgen, dass mehr gezielte flüssige Informationsflüsse initiiert werden, und *verbessert* somit den Informationsfluss.

**Tabelle 19. Einordnung der Technik "FLOW-Mapping"**

**FLOW-Mapping**

| | | | | | |
|---|---|---|---|---|---|
| **Phase** | | ☑ Erheben | ☑ Analysieren | ☑ Verbessern | |
| **Ziel** 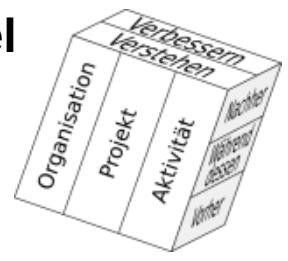 | **Absicht** | ☑ Verstehen | ☑ Verbessern | | |
| | **Zeit** | ☐ Vorher | ☑ Während dessen | ☐ Nachher | |
| | **Umfang** | ☐ Aktivität | ☑ Projekt | ☐ Organisation | |
| **Phasenspezifische Aspekte** | | | | | |
| **Strategie** | **Erhebung** | ☑ Bottom-Up | ☐ Top-Down | | |
| | **Analyse** | ☑ manuelle | ☑ teilautomatische | ☐ automatische Analyse | |
| | **Verbesserung** | O Nebenprodukt<br>O Leichtgewichtig | ⊗ Hauptprodukt<br>O Schwergewichtig | | |
| **Verfahren** | **Erhebung** | ☐ Selber machen<br>☐ Modellableitung | ☐ Beobachten<br>☑ Kommunikationsereignisableitung | ☐ Interview | ☐ Fragebogen |
| | **Analyse** | ☑ Visualisierung | ☐ Simulation | ☐ Mustersuche | |
| | **Verbesserung** | ☐ Aggregatzustand<br>☑ Fluss<br><br>☐ Aktivität | ( O verfestigen<br>( ⊗ Abkürzung<br>O Verzweigung<br>( O Schnittstellenanpassung ⊗ Aktivitätsanpassung) | O verflüssigen )<br>O Umweg<br>O Zusammenführung ) | |

## 4.5 Interface Variationstechnik

### Beschreibung

Die Interface-Variationstechnik ist eine FLOW-Technik, bei der die Informationsflussschnittstelle einer Entwicklungsaktivität geändert wird, um eine Verbesserung zu erzielen. Z.B. können Aktivitäten beschleunigt werden, indem eingehende Informationen flüssig statt bisher fest verarbeitet oder ausgehende Informationen in nur noch ein statt bisher drei Dokumente geschrieben werden.

Eine ausführliche Beschreibung zur Interface-Variationstechnik befindet sich in folgender Veröffentlichung [14]:

> Kurt Schneider, Daniel Lübke: *Systematic Tailoring of Quality Techniques*. World Congress of Software Quality 2005, Munich, Germany, 2005

### Zusammenfassung

Die wesentlichen Schritte der Interface-Variationstechnik sind:

1. Erheben: Aktivität modellieren
    a. Eingehende, ausgehende und steuernde Flüsse und Speicher inkl. Aggregatzustand
2. Analysieren: FLOW-Interface analysieren
3. Verbessern: FLOW-Interface anpassen
    a. Weniger oder mehr eingehende oder ausgehende Flüsse
    b. Andere Aggregatzustände der eingehenden oder ausgehenden Informationen
4. Verbessern: Aktivität an neues Interface anpassen

Die **Interface Variationstechnik** ist eine Verbesserungstechnik, die auch Aktivitäten der Erhebung und Analyse beinhaltet. Mit ihr werden Aktivitäten vor Projektstart gezielt angepasst, indem das FLOW-Interface, d.h. die eingehenden und ausgehenden Flüsse einer FLOW-Aktivität, verändert wird.

**Tabelle 20. Einordnung der Interface Variationstechnik**

**Interface Variationstechnik**

| | | | | | |
|---|---|---|---|---|---|
| **Phase** | | ☑ Erheben | ☑ Analysieren | ☑ Verbessern | |
| **Ziel** 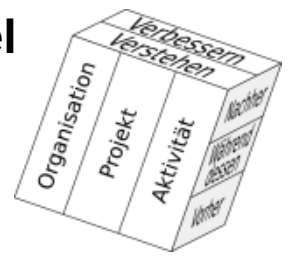  | **Absicht** | ☐ Verstehen | ☑ Verbessern | | |
| | **Zeit** | ☑ Vorher | ☐ Während dessen | ☐ Nachher | |
| | **Umfang** | ☑ Aktivität | ☐ Projekt | ☐ Organisation | |
| **Phasenspezifische Aspekte** | | | | | |
| **Strategie** | **Erhebung** | ☐ Bottom-Up | ☐ Top-Down | | |
| | **Analyse** | ☑ manuelle | ☐ teilautomatische | ☐ automatische Analyse | |
| | **Verbesserung** | O Nebenprodukt<br>O Leichtgewichtig | ⊗ Hauptprodukt<br>O Schwergewichtig | | |
| **Verfahren** | **Erhebung** | ☐ Selber machen<br>☐ Modellableitung | ☐ Beobachten<br>☐ Kommunikationsereignisableitung | ☐ Interview | ☐ Fragebogen |
| | **Analyse** | ☑ Visualisierung | ☐ Simulation | ☐ Mustersuche | |
| | **Verbesserung** | ☐ Aggregatzustand<br>☐ Fluss<br><br>☑ Aktivität | ( O verfestigen<br>( O Abkürzung<br>O Verzweigung<br>( ⊗ Schnittstellenanpassung | O verflüssigen )<br>O Umweg<br>O Zusammenführung )<br>O Aktivitätsanpassung) | |

## 4.6 Integration von SCRUM in das V-Modell XT

### Beschreibung

Die Integrationstechnik beschreibt, wie SCRUM [16] in auf V-Modell XT [21] basierte Projekte integriert werden kann. Um geeignete Integrationspunkte zu finden und um herzuleiten, was bei einer Integration zu beachten ist, werden Informationsflüsse in SCRUM und im V-Modell betrachtet. Das Ziel der Technik ist es, Informationsflüsse in V-Modell-Projekten durch punktuelle Nutzung von SCRUM an geeigneten Stellen zu verbessern. Agile Vorteile wie ein frühzeitig lauffähiges Produkt, regelmäßiges Kundenfeedback, daraus resultierende Risikominimierung, und weniger sowie effizientere Dokumentation sollen dadurch auch V-Modell-Projekten zugänglich gemacht werden.

Abbildung 14 zeigt die Aktivitäten und Informationsflüsse der Integrationstechnik, wie sie Kiesling in seiner Masterarbeit entwickelt hat [8]. Zunächst werden ausgehend von den Beschreibungen des V-Modells und von SCRUM Informationsflüsse bzw. Informationsflussschnittstellen abgeleitet. Diese werden im nächsten Schritt auf geeignete Integrationspunkte untersucht. Dabei ergeben sich Integrationsvarianten und Bedingungen, die es bei einer Integration zu beachten gilt. Die eigentliche Verbesserung findet im letzten Schritt durch die Integration von SCRUM in das V-Modell-XT-Projekt statt.

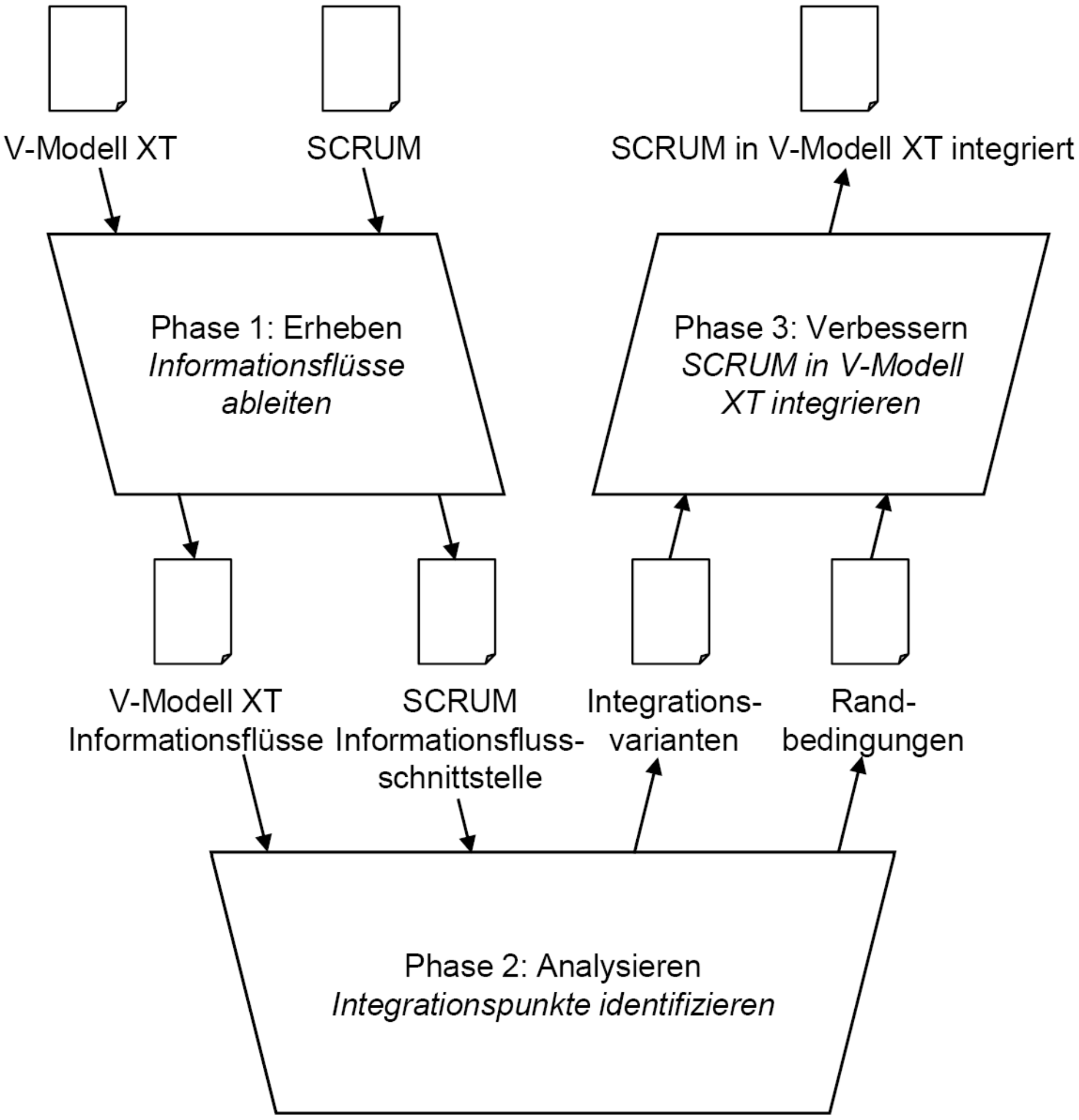

**Abbildung 14. Aktivitäten der SCRUM-Integration**

Die Aktivitäten werden in der Masterarbeit von Kiesling genauer beschrieben:

> Stephan Kiesling, *Integration agiler Entwicklungszyklen in das V-Modell auf Basis von Informationsflüssen*, Masterarbeit, Leibniz Universität Hannover, 2011

### Zusammenfassung

Die Integrationstechnik zeigt, dass sich der FLOW-Ansatz eignet, die beiden Welten der agilen und Dokumenten-zentrierten Softwareentwicklung aus einem einheitlichen Blickwinkel zu betrachten und die jeweiligen Vorteile gezielt zu vereinen. Durch Erhebung, Analyse und schließlich Verbesserung der Informationsflüsse in V-Modell-Projekten bietet die Technik Unterstützung für alle drei Phasen der FLOW-Methode.

Zusammenfassend lässt sich das allgemeine Vorgehen zur Integration von SCRUM in das V-Modell XT wie folgt beschreiben:

**Phase 1: Informationsflüsse herleiten:**

1. Alle V-Modell-Informationsflüsse für einen getailorten V-Modell-Prozess herleiten.

**Phase 2: Integrationspunkte identifizieren:**

2. A: Freie Integration:
   a. Das mit SCRUM zu erstellende Ziel-Produkt festlegen.
   b. Gewünschte Quell-Produkte festlegen.
2. B: Integration entlang von Entscheidungspunkten:
   a. Den mit SCRUM zu erreichenden Ziel-Entscheidungspunkt festlegen
   b. Gewünschten Start-Entscheidungspunkt festlegen
3. Im V-Modell-Informationsfluss alle Zwischenprodukte zwischen Quell- und Ziel-Produkten identifizieren.
4. Alle Produkte identifizieren, die von Zwischenprodukten abhängig sind und weder selbst Zwischenprodukt noch Ziel-Produkt sind. Diese Produkte werden zu weiteren Ziel-Produkten der SCRUM-Integration.

**Phase 3: SCRUM integrieren:**

5. Vorteile gegen Nachteile abwägen.
6. Entsprechend der gewählten Integrationsvariante integrieren.

Tabelle 21 zeigt die Einordnung der Integrationstechnik in die FLOW-Methode mit Hilfe des Templates aus Tabelle 6.

**Tabelle 21. Einordnung der SCRUM-Integrations-Technik**

**SCRUM-Integration**

| | | | | | |
|---|---|---|---|---|---|
| **Phase** | | ☑ Erheben | ☑ Analysieren | ☑ Verbessern | |
| **Ziel** 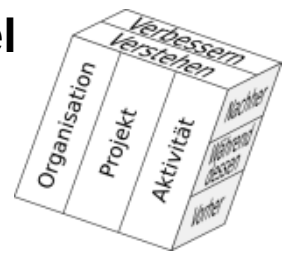 | **Absicht** | ☐ Verstehen | ☑ Verbessern | | |
| | **Zeit** | ☑ Vorher | ☑ Während dessen | ☐ Nachher | |
| | **Umfang** | ☐ Aktivität | ☑ Projekt | ☐ Organisation | |
| **Projektparameter** | | | | | |
| **Vorgehensmodell** | | ☑ Agil: SCRUM | ☑ Prozess: V-Modell XT | | |
| **Verteiltheit** | | ☑ lokal | ☐ verteilt | | |
| **Phasenspezifische Aspekte** | | | | | |
| **Strategie** | **Erhebung** | ☑ Bottom-Up | ☑ Top-Down | | |
| | **Analyse** | ☑ manuelle | ☑ teilautomatische | ☑ automatische Analyse | |
| | **Verbesserung** | ○ Nebenprodukt<br>○ Leichtgewichtig | ○ Hauptprodukt<br>⊗ Schwergewichtig | | |
| **Verfahren** | **Erhebung** | ☐ Selber machen<br>☑ Modellableitung | ☐ Beobachten<br>☐ Kommunikationsereignisableitung | ☐ Interview | ☐ Fragebogen |
| | **Analyse** | ☑ Visualisierung | ☐ Simulation | ☑ Mustersuche | |
| | **Verbesserung** | ☑ Aggregatzustand<br>☑ Fluss<br><br>☑ Aktivität | ( ○ verfestigen<br>( ⊗ Abkürzung<br>○ Verzweigung<br>( ○ Schnittstellenanpassung ⊗ Aktivitätsanpassung) | ⊗ verflüssigen )<br>○ Umweg<br>⊗ Zusammenführung ) | |

## 4.7 FLOW-Muster

### Beschreibung

Mit Hilfe von FLOW-Mustern werden typische Informationsflusskonstellationen der Softwareentwicklung beschrieben. Dies können sowohl problematische Informationsflüsse, die es zu vermeiden gilt, als auch positive Informationsflüsse, die es nachzuahmen gilt, sein. Um FLOW-Muster in ein einheitliches, wiederverwendbares und nützliches Format zu bringen, wurde in der Masterarbeit von Ge eine FLOW-Muster-Schablone entwickelt:

> Xiaoxuan Ge, FLOW Patterns: Beschreibung und Diskussion von Informationsflussmustern in der Softwareentwicklung (PDF – 2,2MB), Masterarbeit, Leibniz Universität Hannover, 2008

Die Schablone hilft beim Formulieren neuer Muster. Sie enthält u.a. folgende Informationen:

- Lässt sich das Muster mit einem FLOW-Modell charakterisieren?
- Welche Struktur hat das Muster?
- Zu welcher Mustergruppe gehört es?
- Hat es positive oder negative Auswirkungen auf den Informationsfluss in der Softwareentwicklung?
- Gibt es verwandte Muster?
- Gibt es konkrete Beispiele?

Darüber hinaus wurde in der Masterarbeit von Ge ein Katalog mit bis dato bekannten Mustern zusammengestellt:

> Xiaoxuan Ge, FLOW-Patterns-Katalog (PDF – 2,7MB), aus Masterarbeit, Leibniz Universität Hannover, 2008

Dieser dient als Basis für die Umsetzung der FLOW-Muster-Technik. Er wird sowohl für die Phase 2: Analysieren als auch für die Phase 3: Verbessern im FLOW-Verbesserungsprozess benötigt. Doch bevor eine Muster-Analyse mit Hilfe des Katalogs stattfinden kann, müssen die zu analysierenden Informationsflüsse erhoben werden (Phase 1). Die Erhebung der Informationsflüsse für die Musteranalyse ist nicht Teil der FLOW-Muster-Technik. Die Erhebung muss vorher stattgefunden haben. Dies kann zum Beispiel mit der Elicitations-Technik geschehen (vgl. Kapitel 4.1).

Im Folgenden werden einige Beispiele von FLOW-Mustern genannt.

- Stille Post
- Totes Dokument
- Osmose
- Leichtgewichtige Dokumentation
- Bürokratie

Details und weitere Muster finden sich in der Masterarbeit von Ge.

### *Phase 2: Analysieren*

**Mustersuche bzw. Pattern-Matching:** Die Analyse von Informationsflussmodellen geschieht bei der FLOW-Muster-Technik mit Hilfe von Mustersuche. Es werden in einem vorher erhobenen FLOW-Modell Muster gesucht. Dies kann auf Basis des charakteristischen Modells des Musters geschehen oder anhand anderer Aspekte aus der Muster-Beschreibung. Bei einer Übereinstimmung hat man entweder einen Informationsfluss mit positiven oder negativen Auswirkungen gefunden, je nach Eigenschaft des Musters. Aus der Musterbeschreibung lässt sich dann auch entnehmen, welche Auswirkungen und Folgen (positive oder negative) das Vorhandensein des Musters haben kann. So liefert die Musteranalyse auch Hinweise, an welchen Stellen im Informationsfluss noch weiter geforscht werden sollte, z.B. nach Problemursachen.

### *Phase 3: Verbessern*

**Gezielter Einsatz pos. und gezieltes Vermeiden neg. Muster:** Um den Fluss zu verbessern sind negative Muster zu vermeiden oder positive Muster an kritischen Stellen des Projekts gezielt einzusetzen.

## Zusammenfassung

Die Technik der **FLOW-Muster** bedient sich Verfahren der Visualisierung und der Mustersuche. Zusammenfassend kann die FLOW-Muster-Technik wie folgt eingesetzt werden:

**Mustersuche:**

- In den mit Hilfe einer anderen FLOW-Technik erhobenen Informationsflussmodellen werden FLOW-Muster anhand ihres charakteristischen Modells gesucht.
- Identifizierte Muster können in einer gegebenen Situation positive oder negative Auswirkungen haben.

**Mustersetzung:**

- Muster mit negativen Auswirkungen sollten durch positive Muster ersetzt werden oder durch andere Informationsflussveränderungen
- Muster mit positiven Auswirkungen sollten beibehalten werden.
- Neutrale Stellen können durch den Einsatz von Mustern mit positiven Auswirkungen verbessert werden.

Tabelle 22 zeigt die Einordnung der Muster-Technik in die FLOW-Methode mit Hilfe des Templates aus Tabelle 6.

**Tabelle 22. Einordnung der Technik "FLOW-Muster"**

**FLOW-Muster**

| | | | | |
|---|---|---|---|---|
| **Phase** | | ☐ Erheben | ☑ Analysieren | ☑ Verbessern |
| **Ziel** 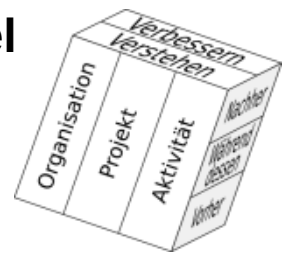 | **Absicht** | ☑ Verstehen | ☑ Verbessern | |
| | **Zeit** | ☐ Vorher | ☑ Während dessen | ☐ Nachher |
| | **Umfang** | ☐ Aktivität | ☑ Projekt | ☐ Organisation |
| **Phasenspezifische Aspekte** | | | | |
| **Strategie** | **Analyse** | ☑ manuelle | ☐ teilautomatische | ☐ automatische Analyse |
| | **Verbesserung** | Musterabhängig | | |
| **Verfahren** | **Analyse** | ☑ Visualisierung | ☐ Simulation | ☑ Mustersuche |
| | **Verbesserung** | Musterabhängig | | |

# 4.8 Erfahrungsverfestigung

## Beschreibung

LIDs steht für Light-Weight Documentation of Experiences - die leichtgewichtige Dokumentation von Erfahrungen [10]. Sie beschreibt eine Technik, mit der man ohne großen Aufwand die vielen am Ende eines Projekts vorhandenen flüssigen Erfahrungen verfestigen kann, sodass sie in späteren gleichartigen Projekten wiederverwendet werden und so zukünftige Projekte verbessern können.

Grundlage der Methode ist ein LID-Template, das durch seine Struktur die Erfahrungserhebung anleitet. Am Ende eines Projekts kommen alle beteiligten Mitarbeiter in einer LID-Sitzung zusammen und lassen das Projekt nochmal Revue passieren, um sich an wichtige Projektsituationen zu erinnern und Erfahrungen dazu festzuhalten. Ein Moderator führt anhand der Struktur und der Anweisungen im LID-Template die Diskussion und schreibt die Erfahrungen der Mitarbeiter in das Dokument. Eine LID-Sitzung dauert zwischen 1,5 und 2 Stunden und ist damit eine einfache und ressourcenschonende Methode zur Erfahrungserhebung.

Eine ausführliche Beschreibung zur Erfahrungsverfestigung mit LIDs befindet sich in folgender Veröffentlichung [10]:

Kurt Schneider: *LIDs: A Light-Weight Approach to Experience Elicitation and Reuse*. PROFES 2000, Oulu, Finland, 2000

## Zusammenfassung

Die **Erfahrungsverfestigung** ist eine *leichtgewichtige* Technik zur *Verfestigung* von in einem *Projekt* gesammelten Erfahrungen nach Beendigung des Projekts. Die festen Erfahrungen können in zukünftigen Projekten genutzt werden und verbessern somit die Informationsflüsse der gesamten Organisation.

Die wesentlichen Schritte der LID-Technik sind:

**Erfahrungsdokumentation:**

- Nach Abschluss eines Projekts alle (relevanten) Projektbeteiligten in einem LID-Meeting zusammen bringen.
- Während des Meetings füllt ein Moderator das LID mit Hilfe eines Templates aus.
- Dazu stellt er den Anwesenden die im Template formulierten Fragen und notiert die Antworten
    - Dabei kommt es nicht auf gute Formulierung und fehlerfreie Schreibweise an
    - Möglichst alle Erfahrungen notieren
    - Kontext und Schlussfolgerung sind wichtig und sollten daher immer notiert werden
    - Auch die Emotion (War das gut oder schlecht?) sollte notiert werden, wenn sie nicht aus dem Kontext hervorgeht. Ggf. muss der Moderator nachfragen.
    - Wichtige Dokumente und andere Informationsquellen, die genannt werden, sollten während der LID-Erstellung schnell unterstrichen werden (Diese werden später verlinkt, siehe unten)
- Nach dem LID-Meeting werden die erwähnten Dokumente im LID verlinkt.
- Ggf. wird das LID für die spätere Auswertung anonymisiert.

**Erfahrungsnutzung:**

- Vor Projektbeginn oder in den frühen Phasen eines Projekts alle LIDs von ähnlichen Projekten durchgehen und die dort genannten Erfahrungen im aktuellen Projekt berücksichtigen.

Tabelle 23 zeigt die Einordnung der LID-Technik in die FLOW-Methode mit Hilfe des Templates aus Tabelle 6.

**Tabelle 23. Einordnung der Technik "Erfahrungsverfestigung"**

**Erfahrungsverfestigung**

| | | | | |
|---|---|---|---|---|
| **Phase** | | ☐ Erheben | ☐ Analysieren | ☑ Verbessern |
| **Ziel** 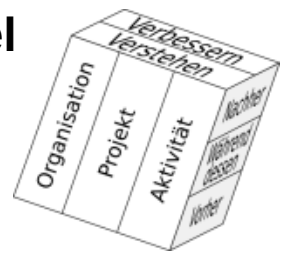 | **Absicht** | ☐ Verstehen | ☑ Verbessern | |
| | **Zeit** | ☐ Vorher | ☑ Während dessen | ☐ Nachher |
| | **Umfang** | ☐ Aktivität | ☐ Projekt | ☑ Organisation |
| **Phasenspezifische Aspekte** | | | | |
| **Strategie** | | ☐ By-Product | ☑ Leichtgewichtig | |
| **Verfahren** | | ☑ Aggregatzustand<br>☐ Fluss<br><br>☐ Aktivität | ( ⊗ verfestigen<br>( O Abkürzung<br>O Verzweigung<br>( O Schnittstellenanpassung | O verflüssigen )<br>O Umweg<br>O Zusammenführung )<br>O Aktivitätsanpassung ) |

# 4.9 Schnelle Anforderungsverfestigung

## Beschreibung

Die schnelle Anforderungsverfestigung ist eine Technik, bei der möglichst viele und qualitätsgesicherte Anforderungen in nur einem Kunden-Interview erhoben und dokumentiert werden. Mit Hilfe eines Werkzeugs, welches es ermöglicht, mit dem Kunden Use Cases und GUI-Mockups zu erstellen, können unter Umständen zeitaufwendige Folgeinterview eingespart werden. Durch Erstellung und Präsentation von Oberflächen-Prototypen inkl. einfacher Maskenabläufe direkt im ersten Interview kann bereits erstes Kundenfeedback berücksichtigt und so die Qualität der dokumentierten Anforderungen verbessert werden.

Eine ausführliche Beschreibung zur Anforderungsverfestigung mit FastFeedback befindet sich in der Bachelorarbeit von Carl Volhard und der folgenden Veröffentlichung von Kurt Schneider und Thao Nguyen [15]:

> Carl Volhard, Werkzeug zur Unterstützung von Interviews in der Prozessmodellierung, Bachelorarbeit, Leibniz Universität Hannover, 2006
>
> Kurt Schneider, Thao Nguyen: *FastFeedback – schnelle Anforderungserhebung mit hoher Ausbeute*. Software Quality Conference (SQC 2007), SQS, Düsseldorf, 2007

Das Werkzeug FastFeedback wurde später in der Diplomarbeit von Melanie Hennemann evaluiert:

> Melanie Hennemann, Ein kontrolliertes Experiment über die Auswirkung von Feedback-Werkzeugen auf die Anforderungserhebung, Diplomarbeit, Leibniz Universität Hannover, 2010

## Zusammenfassung

Die wesentlichen Schritte der schnellen Anforderungsverfestigung mit FastFeedback sind:

**Erstes Kundengespräch:**

- Dokumentation der funktionalen Anforderungen in Use Cases im FastFeedback-Werkzeug
- Erstellung von Oberflächenskizzen mit FastFeeback
- Verknüpfung von Use-Case-Schritten mit Oberflächenskizzen mit FastFeedback
- Abspielen und Präsentation der Maskenabläufe mit FastFeedback
- Berücksichtigung des dabei anfallenden Kundenfeedbacks

Die **schnelle Anforderungsverfestigung** ist eine *Nebenprodukt*-Technik zur *Verfestigung* von Anforderungen. Sie unterstützt die *Aktivität* der Anforderungserhebung *während* der Projektlaufzeit. Die Verfestigung der Anforderungen als Nebenprodukt beschleunigt den Anforderungsprozess und führt zu qualitativ hochwertigeren Anforderungen und *verbessert* somit die Softwareentwicklung. Tabelle 24 ordnet die FLOW-Technik der schnellen Anforderungsverfestigung in die FLOW-Methode ein.

**Tabelle 24. Einordnung der Technik "Anforderungsverfestigung"**

**Schnelle Anforderungsverfestigung**

| | | | | |
|---|---|---|---|---|
| **Phase** | | ☐ Erheben | ☐ Analysieren | ☑ Verbessern |
| **Ziel** 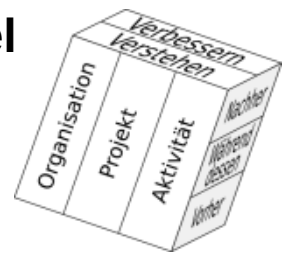  | **Absicht** | ☐ Verstehen | ☑ Verbessern | |
| | **Zeit** | ☐ Vorher | ☑ Während dessen | ☐ Nachher |
| | **Umfang** | ☑ Aktivität | ☐ Projekt | ☐ Organisation |
| **Phasenspezifische Aspekte** | | | | |
| **Strategie** | | ☑ By-Product | ☑ Leichtgewichtig | |
| **Verfahren** | | ☑ Aggregatzustand | ( ⊗ verfestigen | O verflüssigen ) |
| | | ☐ Fluss | ( O Abkürzung | O Umweg |
| | | | O Verzweigung | O Zusammenführung ) |
| | | ☐ Aktivität | ( O Schnittstellenanpassung | O Aktivitätsanpassung ) |

# 4.10 Prototyp-Demo-Verfestigung

## Beschreibung

Die Technik zur Verfestigung des Wissens, welches bei einer Demonstration eines Prototyps zwischen Entwicklern ausgetauscht wird, hilft, dieses wertvolle Wissen langfristig zugreifbar zu machen. Dazu wird mit Hilfe eines speziellen Werkzeugs während der Prototypendemonstration folgendes aufgezeichnet:

- Ein Screencast des Demonstrations-PCs. Dabei wird sowohl die Oberfläche des Prototyps als auch der während der Demonstration gezeigte Quellcode aufgezeichnet.

- Ein Audiomitschnitt der mündlichen Erläuterungen während der Demonstration.
- Ein Index des gerade gezeigten Quellcodeausschnitts über die integrierte Entwicklungsumgebung (IDE).

Diese Informationen können später von Entwicklern, die nicht an der ursprünglichen Prototypendemonstration teilgenommen haben, genutzt werden, um sich in den Quellcode des Prototyps einzuarbeiten und diesen besser zu verstehen.

Eine ausführliche Beschreibung der Prototyp-Verfestigungs-Technik findet sich in folgendem Paper:

> Kurt Schneider: *Prototypes as Assets, not Toys. Why and How to Extract Knowledge from Prototypes*, 18th International Conference on Software Engineering (ICSE-18), 1996.

## Zusammenfassung

Die **Prototyp-Demo-Verfestigung** ist eine *leichtgewichtige Nebenprodukt*-Technik zur *Verfestigung* von Informationen, die bei einer Demonstration eines Prototyps zwischen Entwicklern ausgetauscht werden. Die so verfestigten Informationen erleichtern zum Beispiel neuen Entwicklern den Einstieg in das Projekt.

**Tabelle 25. Einordnung der Technik "Prototyp-Demo-Verfestigung"**

**Prototyp-Demo-Verfestigung**

| | | | | |
|---|---|---|---|---|
| **Phase** | | ☐ Erheben | ☐ Analysieren | ☑ Verbessern |
| **Ziel** 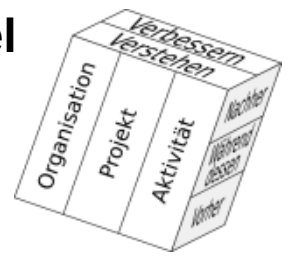 | **Absicht** | ☐ Verstehen | ☑ Verbessern | |
| | **Zeit** | ☐ Vorher | ☑ Während dessen | ☐ Nachher |
| | **Umfang** | ☑ Aktivität | ☐ Projekt | ☐ Organisation |
| **Phasenspezifische Aspekte** | | | | |
| **Strategie** | | ☑ By-Product | ☑ Leichtgewichtig | |
| **Verfahren** | | ☑ Aggregatzustand | ( ⊗ verfestigen | O verflüssigen ) |
| | | ☐ Fluss | ( O Abkürzung | O Umweg |
| | | | O Verzweigung | O Zusammenführung ) |
| | | ☐ Aktivität | ( O Schnittstellenanpassung | O Aktivitätsanpassung ) |

# 4.11 Verfestigung als Nebenprodukt

## Beschreibung

Fehlende Dokumentation ist ein typisches Problem der Softwareentwicklung, insbesondere in global verteilten Projekten [18]. Die FLOW-Technik zur Verfestigung von Kommunikation als Nebenprodukt hilft, wichtige Informationen, z.B. aus verteilten Meetings, so zu dokumentieren, dass sie später leicht wieder gefunden werden können. Um die Teilnehmer des Meetings bei der Dokumentationserstellung möglichst wenig bei ihrer eigentlichen Arbeit zu stören und trotzdem nützliche, gut strukturierte Dokumentation zu bekommen, wird der Nebenprodukt-Ansatz von Schneider [12] verwendet. Dieser sieht vor, dass Informationen während einer Tätigkeit automatisch aufgezeichnet werden und dabei so viele Zusatzinformationen wie möglich, aber ohne dabei die eigentliche Aufgabe zu stören, für die Erstellung eines Index mit aufgezeichnet werden. Falls die Aufzeichnung noch weiter strukturiert und indexiert werden soll, soll dies in einem separaten Arbeitsschritt passieren. Der Index ist notwendig, um später schneller die relevanten Informationen in einer potenziell sehr großen Datenmenge (z.B. Audiomitschnitt eines drei stündigen Meetings) wiederfinden zu können.

Abbildung 15 zeigt die wesentlichen Informationsflüsse der Verfestigungstechnik. Während ihrer regulären Arbeit kommunizieren die Entwickler. Ein speziell auf eine bestimmte Kommunikationsaktivität ausgelegtes Verfestigungswerkzeug erstellt eine Dokumentation inklusive Index nach [12]. Diese Dokumentation kann zu einem späteren Zeitpunkt von den Entwicklern genutzt werden, um wichtige Informationen wiederzufinden. Der Index hilft dabei, diese Informationen schneller wiederzufinden.

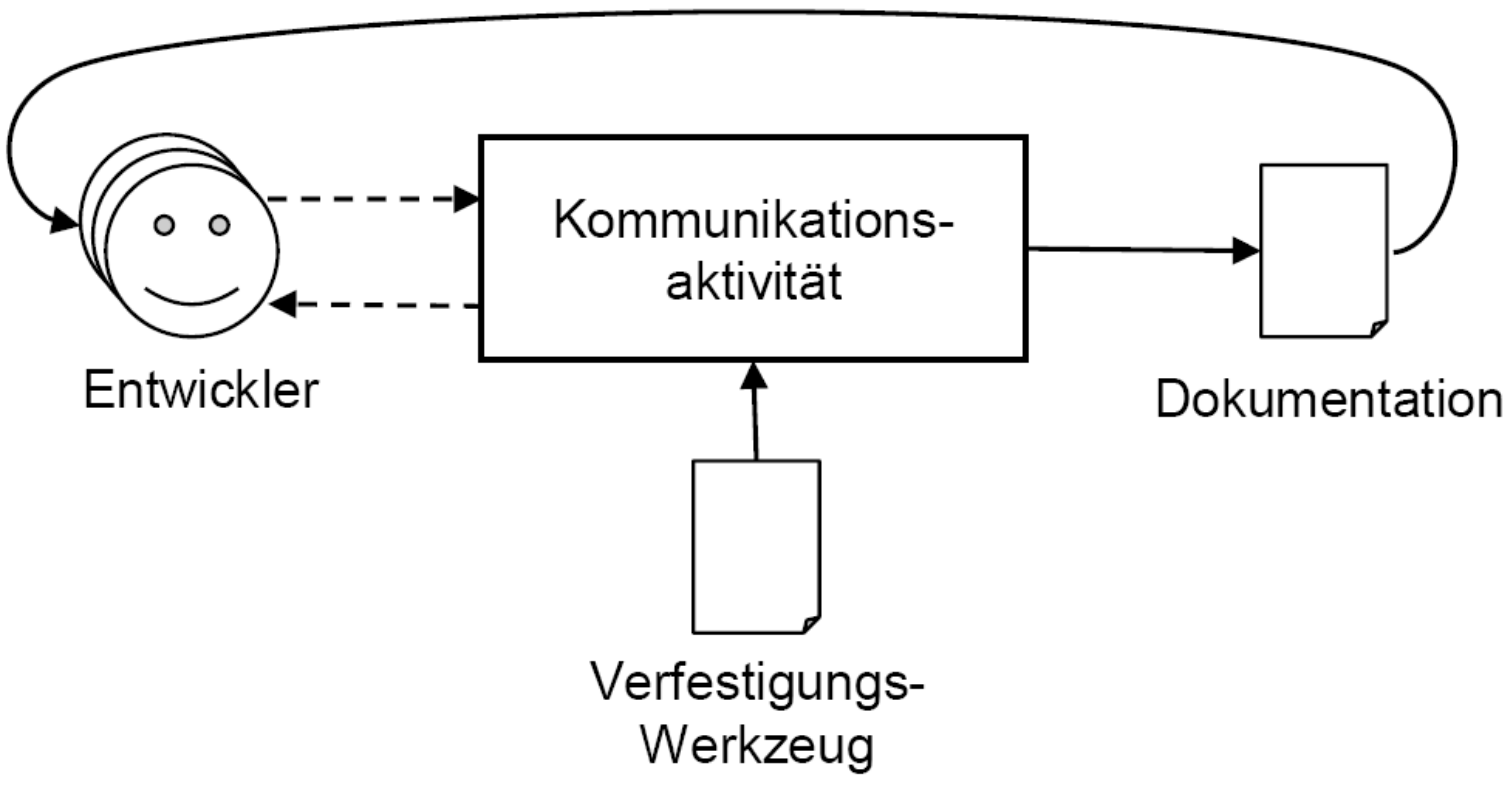

**Abbildung 15. Informationsflüsse der Verfestigung-als-Nebenprodukt-Technik**

Song [17] hat in ihrer Bachelorarbeit ein Werkzeug entwickelt, welches die Nebenprodukttechnik für Skype-basierte Meetings nutzbar macht. In dieser Arbeit wurde auch die Nützlichkeit eines Index für Audiomitschnitte evaluiert.

Hui Song, Erstellung eines Werkzeugs zur Unterstützung der Dokumentation von Skype-basierten Meetings, Bachelorarbeit, Leibniz Universität Hannover, 2011

## Zusammenfassung

Ziel der Technik der Dokumentation als Nebenprodukt ist die Verbesserung des Informationsflusses durch beiläufige Verfestigung flüssiger Informationsflüsse und zusätzlicher Strukturierung dieser verfestigten Informationen durch einen Index. Die Technik wird während eines Meetings oder anderen Kommunikationsaktivitäten eingesetzt. Es wurde ein Werkzeug geschaffen, dass die Verfestigungstechnik für Skype-Telefonkonferenzen umsetzt [17]. Damit ist es möglich, strukturierte Dokumentation als Nebenprodukt einer Telefonkonferenz zu erhalten. Die während einer Konferenz verschickten Textnachrichten werden automatisch zur Erstellung eines Index genutzt. Weiterhin kann mit Hilfe des Werkzeugs der Index während oder nach dem Meeting bearbeitet werden.

Tabelle 26 zeigt die Einordnung der Verfestigungstechnik in die FLOW-Methode mit Hilfe des Templates aus Tabelle 6.

**Tabelle 26. Einordnung der Technik "Verfestigung als Nebenprodukt"**

| | | Verfestigung als Nebenprodukt | | |
|---|---|---|---|---|
| **Phase** | | ☐ Erheben | ☐ Analysieren | ☑ Verbessern |
| **Ziel** 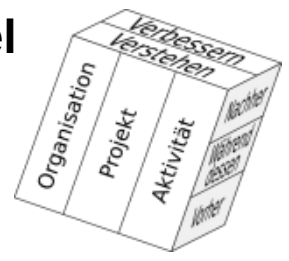 | **Absicht** | ☐ Verstehen | ☑ Verbessern | |
| | **Zeit** | ☐ Vorher | ☑ Während dessen | ☐ Nachher |
| | **Umfang** | ☑ Aktivität | ☐ Projekt | ☐ Organisation |
| **Phasenspezifische Aspekte** | | | | |
| **Verbesserungsstrategie** | | ☑ Nebenprodukt | ☑ Leichtgewichtig | |
| **Verbesserungsverfahren** | | ☑ Aggregatzustand<br>☑ Fluss<br><br>☐ Aktivität | ( ⊗ verfestigen<br>( O Abkürzung<br>⊗ Verzweigung<br>( O Schnittstellen- | O verflüssigen )<br>O Umweg<br>O Zusammenführung )<br>O Aktivitätsanpassung ) |

# A. Anhang: Elicitation Package

Das Elicitation Package besteht aus zwei Teilen: (1) Einem *Fragebogen*, der die Erhebung anleitet und (2) einem *Erhebungsbogen*, der mehrfach, je für eine zentrale Aktivität, ausgefüllt wird.

## A.1 Fragebogen

(siehe die folgenden beiden Seiten)

Interviewer: ___________________________ ID: ______

# FLOW Elicitation Fragebogen

Phase: Erheben Technik: Informationsflüsse elicitieren

| | |
|---|---|
| Name | |
| Kontakt (✉/☎) | |
| Abteilung / Gruppe | |
| Datum | Start: Ende: |

## 1. Allgemein

1. Wie lange sind Sie schon Teil ihrer Abteilung / Gruppe?

2. Ausbildungshintergrund?

3. Professioneller Werdegang?

4. Ihr Arbeitsplatz? (Wieviele im Büro? Wie oft Störungen?)

5. Ihre Hauptaufgabe? Tätigkeitsbezeichnung?

6. Welche Tools, Prozesse, Programmiersprachen benutzen Sie während Ihrer täglichen Arbeit?

## 2. Informationsflüsse

Benutzen Sie die *Erhebungsbögen* für diesen Teil und fassen Sie unten zusammen. Zudem können die folgenden Fragen als Anleitung zum Ausfüllen des Erhebungsbogens genutzt werden.

### *2.1. FLOW-Interface (→ Erhebungsbogen)*

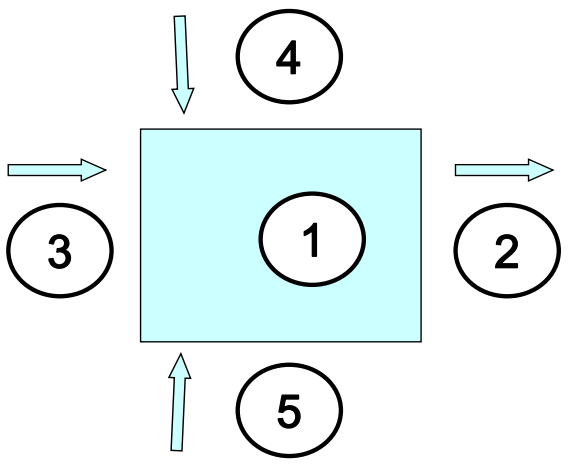

1. Was sind Ihre Hauptaufgaben? → je ein Bogen

2. Wer nutzt Ihre Arbeitsergebnisse? In welcher Form/Medium? → Output

3. Welche Informationen benötigen Sie für Ihre Arbeit? → Input

4. Müssen Sie bei Ihrer Arbeit bestimmten Vorgaben folgen? (mündlich/schriftlich)
   → Steuerung

5. Welche Werkzeuge oder Personen unterstützen die Durchführung Ihrer Arbeit?
   → Unterstützung

### *2.2. Flüssige Information und Erfahrung*

Weitere Fragen, um andere Informationsquellen und –senken aufzudecken:

1. Gibt es andere Stakeholder ihrer Arbeit?

2. Mit wem arbeiten Sie zusammen?

3. Welche Meetings/Gespräche/Abstimmungen sind für die Funktionen wichtig?

4. Wer prüft und wogegen? Was geschieht bei Ablehnung?

5. Wie werden die Ergebnisse festgehalten, geprüft, freigegeben?

6. Wer hat auf dem Gebiet die meisten Erfahrungen? Wie profitieren Sie u. die anderen davon?

7. Wie viel Prozent der Informationen bekommen Sie schriftlich?

8. Wie viel Prozent der Informationen bekommen Sie mündlich?

9. Wie viele Informationen fließen über die Gruppenleiter?

## 3. Allgemein

1. Was funktioniert gut?

2. Wo gibt es Probleme?

## A.2 Erhebungsbogen

FLOW Erhebungsbogen

ID:

**Organisation**
Ausfüller
Datum
Gesprächspartner
Kontext (z.B.Projekt)

**Steuerung**

| Funktion/Person | Inhalt | Fest/flüssig | Form/Medium |
|---|---|---|---|
| | | | |
| | | | |
| | | | |

**von**

| Fkt./ Person | Inhalt | fest/ flüssig | Form/ Medium |
|---|---|---|---|
| | | | |
| | | | |
| | | | |
| | | | |
| | | | |
| | | | |

Funktion
(Profil)
Person
(gibt es noch andere)
Aufgabe
Besonderes

**an**

| Fkt./ Person | Inhalt | fest/ flüssig | Form/ Medium |
|---|---|---|---|
| | | | |
| | | | |
| | | | |
| | | | |
| | | | |
| | | | |

**Unterstützung**

| Tool/Person | Inhalt | Fest/flüssig | Form/Medium |
|---|---|---|---|
| | | | |
| | | | |
| | | | |

Fachgebiet Software Engineering
Leibniz Universität Hannover
DFG Projekt InfoFLOW, 2008 – 2011

02.10.2009
Kurt Schneider & Kai Stapel
www.se.uni-hannover.de/forschung/flow

# Literaturverzeichnis